\documentclass[3p, times]{elsarticle}

\usepackage{exscale,times}
\usepackage[english]{babel}
\usepackage{graphicx}
\usepackage{amsmath}
\usepackage{amsfonts}
\usepackage{amssymb}
\usepackage{hyperref}
\usepackage{color}
\usepackage{bm}
\usepackage{mathrsfs}
\usepackage{multirow}
\usepackage{subfig}
\usepackage{calc}
\usepackage{stfloats}
\usepackage{comment}
\usepackage{algorithm}
\usepackage{algorithmic}
\usepackage[flushleft]{threeparttable}

\newcommand{\tabincell}[2]{\begin{tabular}{@{}#1@{}}#2\end{tabular}}

\graphicspath{{images/}}

\begin{document}
\title{A Message Passing Approach for Multiple Maneuvering Target Tracking}

\author[address1,address2,address3,address4]{Hua Lan}
\author[address1,address2]{Jirong Ma}
\author[address1,address2]{Zengfu Wang}
\author[address1,address2]{Quan Pan}
\author[address5]{Xiong Xu}

\address[address1]{School of Automation, Northwestern Polytechnical University, Xi'an, Shaanxi, 710072, China}
\address[address2]{Key Laboratory of Information Fusion Technology, Ministry of Education, Xi'an, Shaanxi, 710072, China}
\address[address3]{School of Electronic Engineering, Xidian University, Xi'an, Shaanxi, 710071, China.}
\address[address4]{National Lab of Radar Signal Processing, Xi'an, Shaanxi, 710071, China.}
\address[address5]{Southwest China Institute of Electronic Technology, Chengdu, Sichuan, 610036, China.}
%\author{Hua Lan\thanks{ All authors are with the School of Automation, Northwestern Polytechnical University, and the Key Laboratory of Information Fusion Technology, Ministry of Education, Xi'an, Shaanxi Province, 710072, PR China. Hua Lan is also with the School of Electronic Engineering, Xidian University, and the National Lab of Radar Signal Processing, Xi'an, Shaanxi Province, 710071, PR China. This work was supported by the National Natural Science Foundation of China (Grant No.~61873211, 61503305, 61790552).}, Jirong Ma, Zengfu Wang, and Quan Pan}

\journal{Signal Processing}

%\maketitle
\begin{frontmatter}
\begin{abstract}
This paper considers the problem of detecting and tracking multiple maneuvering targets, 
which suffers from the intractable inference of high-dimensional latent variables that include target kinematic state, target visibility state, motion mode-model association, and data association.
A unified message passing algorithm that combines belief propagation~(BP) and mean-field~(MF) approximation is proposed for simplifying the intractable inference.
By assuming conjugate-exponential priors for target kinematic state, target visibility state, and motion mode-model association, 
the MF approximation decouples the joint inference of target kinematic state, target visibility state, motion mode-model association into individual low-dimensional inference, yielding simple message passing update equations.
The BP is exploited to approximate the probabilities of data association events since it is compatible with hard constraints.
Finally, the approximate posterior probability distributions are updated iteratively in a closed-loop manner, which is effective for dealing with the coupling issue between the estimations of target kinematic state and target visibility state and decisions on motion mode-model association and data association.
The performance of the proposed algorithm is demonstrated by comparing with the well-known multiple maneuvering target tracking algorithms, including interacting multiple model joint probabilistic data association, interacting multiple model hypothesis-oriented multiple hypothesis tracker and multiple model generalized labeled multi-Bernoulli.
\end{abstract}

%\begin{IEEEkeywords}
%Maneuvering targets; joint detection and tracking; variational Bayes; belief propagation
%\end{IEEEkeywords}

\end{frontmatter}
\section{Introduction}
\label{sec:introduction}
Exploiting noisy measurements from sensors~(e.g., radar and sonar),
joint target detection and tracking is the process of detecting the existence of targets and estimating their kinematic states.
It plays an essential role in many applications, such as surveillance, traffic control, and navigation~\cite{Bar-Shalom2001}, etc.
Joint detection and tracking of multiple maneuvering targets is particularly challenging due to the following reasons:
(1) Data association uncertainty often occurs since the origin of the measurements is unknown in multiple target tracking with clutter.
(2) Targets of interest are typically non-cooperative. Their motion patterns cannot be modeled with absolute confidence.
(3) The number of targets is unknown and time-varying. Targets may appear or disappear at any time in any place of the area of interest.
(4) A tracker has to handle high-dimensional latent variables that include target kinematic state, target visibility state, motion mode-model association, and data association.
(5) The estimations of target kinematic state and target visibility state are coupled with the decisions on data association and motion mode-model association.

Joint detection and tracking of multiple maneuvering targets requires solving four major problems,
kinematic state estimation, data association decision, target motion mode identification, and target detection.
Each of these problems has its line of research, which has been active for decades in the statistical signal processing society.
Specifically, aiming to infer the kinematic state of targets from noisy measurements,
kinematic state estimation can be solved by Kalman filter for linear Gaussian models, and by extended Kalman filter, unscented Kalman filter and particle filter for nonlinear models~\cite{sarkka2013bayesian}.
Data association, determining which measurements are used to update each track, is addressed by methods such as joint probabilistic data association (JPDA)~\cite{JPDA2009}, multiple hypothesis tracker~(MHT)~\cite{MHT2004}, probabilistic multihypothesis tracker~(PMHT)~\cite{PMHT2005}, and Markov chain Monte Carlo data association~(MCMCDA)~\cite{oh2009markov}.
Maneuvering targets may switch between different motion modes.
Using a bank of different hypothetical motion models that follow finite-state Markov chain,
interacting multiple model~(IMM) estimator~\cite{blom1988interacting} is commonly exploited for maneuvering target tracking.
Comprehensive surveys on maneuvering target tracking and data association can be found in \cite{Li2005Survey} and \cite{pulford2005taxonomy}, respectively.
Typical target detection approaches include M-of-N logic~\cite{Barshon2011}, visibility model-based method~\cite{colegrove1986track},
existence model-based method~\cite{musicki1994integrated}, Hough transform~\cite{evans1994search} and random sample consensus~\cite{niedfeldt2017comparison}.
The random finite set~(RFS)-based multitarget tracking methods, including probability hypothesis density~(PHD) filter~\cite{Maler2007}, multi-Bernoulli filter~\cite{vo2009cardinality}, bypass the complicated data association problem and incorporate the joint target detection and tracking in a Bayesian way.
Gradually, these work has been extended to simultaneously handle two or more problems by combination.
For example, IMMJPDA~\cite{IMMJPDA} and IMMPMHT~\cite{IMMPMHT} are proposed to track multiple maneuvering targets.
The joint integrated probabilistic data association~(JIPDA)~\cite{JIPDA} that integrates modeling of target existence with JPDA,
the PMHT with visibility model (PMHT-v)~\cite{PMHTV}, and the belief propagation~(BP) with existence model~\cite{meyer2017scalable} are used for multiple target joint detection and tracking.
IMMMHT~\cite{IMMMHT}, MMPHD~\cite{IMMPHD, georgescu2012multiple},
and multiple model generalized labeled multi-Bernoulli~(MMGLMB)~\cite{reuter2015}, considered all four above-mentioned problems.

In principle, the problem of multiple maneuvering target tracking and detection~(MMTT) can be formulated in a Bayesian framework and solved by computing the joint probability distribution function~(PDF) of high-dimensional latent variables, including target kinematic state, target visibility state, data association and motion mode-model association.
However, exact computation of this joint PDF is intractable by the fact that the required integrations over continuous latent variables~(i.e., target kinematic state) may not have closed-form analytical solutions, and the marginalization involves summing over all possible configurations of the discrete latent variables~(i.e., target visibility state, data association and motion mode-model association).
Two kinds of approximation methods, mean-field~(MF) approximation~\cite{xing2002generalized} and BP,
%\footnote{Following the convention used in ~\cite{yedidia2005constructing,yedidia2003understanding}, we used the name BP also for loopy BP. Specially, BP is used to find the exact marginal of a certain of PDF on a cycle-free factor graph. When the loops exist in a factor graph, the loopy BP is used to find the approximate marginal.}~\cite{yedidia2003understanding}},
are commonly used to solve the high-dimensional inference problem.
In the MF approximation, the intractable joint PDF of high-dimensional latent variables is approximated by tractable fully factorized PDF, and the Kullback-Leibler~(KL) divergence between the approximate PDF and the true PDF is minimized.
BP devotes to find an exact or approximate marginal distribution.
Both MF and BP can be implemented in an iterative way, such as message passing~(MP).
As stated in~\cite{riegler2012merging}, MF has the virtues of convergent implementation and simple MP update rules for conjugate-exponential models. However, MF is not compatible with hard constraints~\cite{yedidia2005constructing}.
%\footnote{\textcolor{red}{In a factor graph}, it is called a ``hard constraint'' if a factor $f_a(\bm{x}_a)=0$ for some $\bm{x}_a$.}.
BP yields a good approximation of the marginal distribution if the factor graph representing the joint distribution has no short cycles.
Unlike MF, BP is compatible with hard constraints.
However, it may have high complexity.
Riegler \emph{et al.} combined BP and MF approximation in a joint MP approach~\cite{riegler2012merging}.

Recently, BP has been attracting much attention from the target tracking society.
Williams and Lau~\cite{williams2014approximate} presented a graphical model formulation of data association and approximated the marginal association probabilities based on BP.
They proved the convergence of BP for the data association problem and showed that the computational complexity of BP is linear
in the number of targets and measurements.
In~\cite{Williams206Multiple}, they further extended their work to multiple scan data association problem,
for which a convex free energy was constructed and optimized using a primal-dual coordinate ascent method.
Meyer \emph{et al.}~\cite{meyer2017scalable} addressed the problem of multisensor-multitarget joint detection-estimation problem,
where the statistical structure of joint latent variables including target kinematic states, target existence state, and data association was described by a factor graph, and the corresponding joint detection-estimation problem was solved by loopy BP~(LBP).
This work was further extended to unknown and time-varying parameters which were assumed to follow Markov chain models~\cite{soldi2019self},
and LBP was adopted to calculate the marginal posterior distributions of the targets and model parameters.
A comprehensive survey on BP for multitarget tracking can be found in~\cite{meyer2018message}.

As a specific type of variational Bayes whereas the approximated PDF is assumed fully factorized,
MF is also widely used in adaptive state estimation and target tracking problems.
The MF approximation for adaptive Kalman filtering with unknown measurement noise covariance was presented in~\cite{sarkka2009recursive},
which was further extended to both unknown process noise covariance and measurement noise covariance~\cite{Huang2017TAC}, and nonlinear adaptive filtering~\cite{ozkan2013marginalized, yu2019nonlinear}.
Ma \emph{et al.}~\cite{ma2018multiple} considered the multiple model state estimation problem, and approximated the joint state estimation and model identification through MF approximation.
To solve the data association problem in multitarget tracking, L\'{a}zaro-Gredilla \textit{et al.}~\cite{Lazaro2012} introduced a mixture of Gaussian processes of which hyperparameters were learned based on MF.

Few work considered combined BP and MF for multitarget joint detection and tracking.
Turner \emph{et al.}~\cite{NIPS2014} proposed a probabilistic tracking algorithm that integrates state estimation, data association, and track management, whereas the joint PDFs of latent variables are fully factorized based on MF and the data association is approximated by BP. Lan \emph{et al.}~\cite{Lan2019Joint} extended the work of~\cite{NIPS2014} to multipath target detection and tracking whereas one target may produce multiple resolved measurements via different propagation paths.
Lau \emph{et al.}~\cite{Lau2016} presented a structured MF approximation that considered the dependence between target kinematic state and target existence state and approximated data association by BP.
However, to our best knowledge, no one has considered MMTT by using the combined BP-MF approximation.

This work considers the high-dimensional inference arising from MMTT, and provides a derivation of combined BP-MF MP approach to joint estimation of target kinematic state and target visibility state, and decisions on motion mode-model associatioin and data association.
Based upon the factor graph corresponding to a factorization of the joint PDF of the latent variables and a choice for a separation of this factorization into BP and MF factors, we use MF to deal with the target kinematic state estimation, visibility state estimation and motion mode-model association due to its simple MP update rules for conjugate-exponential models, and use LBP to solve the data association
with the one-to-one frame (hard)~constraints. The approximate posterior PDFs are updated iteratively in a closed-loop manner, 
which is effective for dealing with the coupling issue between the estimations on target kinematic state and target visibility state
and decisions on motion mode-model association and data association.
The performance of the proposed algorithm, which is referred as MP-MMTT, is demonstrated by comparing with the well-known multiple maneuvering target tracking algorithms, including IMMJPDA, IMM hypothesis-oriented MHT~(IMMHMHT), and MMGLMB.

The rest of the paper is organized as follows.
The problem formulation of MMTT is described in Section~\ref{sec:ProblemFormulation}.
The approximate posterior PDFs, i.e., \emph{beliefs}, of each latent variables are derived via MP framework in Section~\ref{sec:solution}.
The simulation analysis and conclusion are given in Section~\ref{sec:simulation} and Section~\ref{sec:conclusion}, respectively.

\section{Problem Formulation}\label{sec:ProblemFormulation}
This paper considers MMTT in the presence of clutter.
The detection probability of targets is assumed to be less than one.
In this section, we first present the models of target and measurement, and then introduce the problems of data association and target motion mode-model association.
At last, we discuss MMTT in the Bayesian framework.

\subsection{Modeling of Target and Measurement}
Like \cite{Li2005Survey}, we use \emph{mode}, denoted by $\tau$, to refer to the true and unknown pattern of target motion,
and use \emph{model}, denoted by $m$, to describe the motion mode of a target mathematically.
Note that one motion mode of a target can be represented by one or more models, especially when the target is maneuvering.
\emph{Multiple-model approach}, which assumes a set of models as possible candidates of the true mode in operation at the time,
is a mainstream method for maneuvering target tracking~\cite{Li2005Survey}.
In this paper, we assume that:
(1) The true mode of a target is time-variant;
(2) The mode space of a target at any time $k$ is time-invariant and the same as the assumed model set.
The motion mode-model association event $\tau_k^i = m_k^i \in \left \{1, \ldots, N_M \right \}$ with $N_M$ being the known number of models,
denotes that target $i$ moves according to the $m$th model at time $k$.
The model sequence $\left \{m_1^i, \ldots, m_k^i \right \}$ is a Markov chain with initial probability $\pi_{i, m} = \text{Pr} \left[m_0^i = 1, \ldots, m_0^i = N_M \right]$ and transition probability $T_{i,m}(\tau_1, \tau_2) = \text{Pr}(m_{k}^i = {\tau_2}|m_{k-1}^i = {\tau_1})$ from $\tau_1$ to $\tau_2$.
By these assumptions, we will use $m_k^i$ to denote both the mode and the model of the $i$th target at time $k$.
For a Markov jump linear system, the kinematic state of the $i$th target via the $m$th model follows the equation~\cite{IMMJPDA}:
\begin{equation}\label{state}
x_{k}^i = F_k(m_{k}^i)x_{k-1}^i + w_{k}(m_{k}^i),
\end{equation}
where $x_k^i \in \mathbb{R}^{n_x}$ is the $i$th target kinematic state with $n_x$ being the dimension of target kinematic state.
The model-dependent kinematic state transition function $F_k(m_{k}^i)$ is assumed to be known.
Process noise $w_{k}(m_{k}^i)$ is assumed to be a zero-mean white Gaussian process with covariance matrix $Q_k(m_{k}^i)$.
The initial target kinematic state under each model $m$ are assumed to be Gaussian random variables with mean $\bar x_{0}^{i, m}$ and covariance matrix $P_{0}^{i, m}$.
Denote the joint kinematic states of all targets at time $k$ by $X_k = \left \{x_{k}^1, \ldots, x_k^i, \ldots, x_ k^{N_T} \right \}$,
where $N_T$ is the maximum number of potential targets~(tracks).

To perform target detection~(or track maintenance), a tracker needs the abilities to initialize a new track for a newborn target,
and to terminate the tracks when the corresponding targets disappear.
There are two common models, the existence-based model~\cite{Lau2016} and the visibility-based model~\cite{NIPS2014},
for carrying out target detection in a probabilistic way.
The former represents the target kinematic state as a conditional distribution on target existence state;
that is, there are two different kinds of PDFs for target kinematic state, non-existing target kinematic state PDF and existing target kinematic state PDF.
The latter assumes that the target kinematic state is conditionally independent of target visibility state given data association.
Roughly speaking, the existence-based model, which is used in JIPDA~\cite{JIPDA}, random finite sets based algorithms~\cite{2014LMBF}, etc., is more appropriate for recursive processing whereas the joint PDFs of target kinematic state and target existence state are updated with time.
The visibility-based model is often used for batch processing algorithms, such as PMHT-v~\cite{PMHTV}, variational Bayes tracker~(VBT)~\cite{NIPS2014}, etc.
In this paper, we adopt the visibility-based model.
Define the binary variable $e_{k}^i \in \{0, 1\}$ to represent the visibility~(detection) state of target $i$;
that is, target $i$ is visible at time $k$ if $e_{k}^i = 1$, otherwise target $i$ is invisible.
The evolution of the visibility state $e_{k}^i \in \{0, 1\}$ of target $i$ is modeled as a two-state first-order Markov process with initial probability $\pi_{i, e} = \text{Pr}[e_0^i=0, e_0^i = 1]$ and transition probability $T_{i, e}(\xi_1, \xi_2) = \text{Pr}(e_{k}^i = {\xi_2}|e_{k-1}^i = {\xi_1})$ from $\xi_1$ to $\xi_2$. The decision of target detection (or track management) is made based on its visibility probability $p(e_{k}^ i =1)$, i.e.,
a track is confirmed if its visibility probability is greater than a threshold $\delta_{c}$,
and is terminated when its visibility probability is less than a threshold $\delta_{d}$.

Denote the set of all measurements at time $k$ by $Y_k =  \{y_{k}^1, \ldots, y_{k}^j, \ldots, y_{k}^{N_{k, E}} \}$ with $N_{k, E}$ being the number of measurements at time $k$.
Let $y_k^j \in \mathbb{R}^{n_y}$, $j = 1, \ldots, N_{k, E}$, be the $j$th measurement with $n_y$ being the dimension of the measurement.
Each measurement $y_k^j$ may originate from either a target or clutter.
The measurement originated from clutter is uniformly distributed within the volume of the area of interest $V_G$,
and the number of clutter is assumed to be Poisson distributed \cite{Bar-Shalom2001} with intensity $\lambda V_G$,
where $\lambda$ is the clutter density.
The measurement $y_k^j$ originated from target $i$ is measured according to the measurement model $h_k: \mathbb{R}^{n_x} \rightarrow \mathbb{R}^{n_y}$ with a detection probability $p_d^i$, that is, 
\begin{equation}\label{measurement}
y_k^j = h_{k}(x_{k}^i, m_{k}^i) + v_k(m_{k}^i), %\;  \text{if $y_k^j$ is from target $i$}
\end{equation}
where $v_k(m_k^i) \sim \mathcal{N}(0, R^m_k)$ is assumed to be zero-mean white Gaussian measurement noise with covariance matrix $R^m_k$.
Here, $w_{k}(m_k^i)$, $v_k(m_k^i)$ and $x_{0}^{i,m}$ are assumed to be mutually independent.
For simplicity, here and thereafter, we denote $F_k(m_k^i)$, $Q_k(m_k^i)$, $R_k(m_k^i)$, $h_k(x_k^i, m_k^i)$ as $F_k^{i, m}$, $Q_k^{i, m}$, $R_k^{i, m}$, $h_k^m(x_k^i)$, respectively.
For two functions $f$ and $g$, define $f \overset{c}{=} g$ if $f = g + c$, where $c$ is an additive constant;
$f(x) \propto g(x)$ means that $f(x)$ is equal to $g(x)$ up to a proportionality constant;
$\left\langle f(x) \right\rangle_{g} = \int_x f(x) g(x) d_x$ denotes the expectation of $f(x)$ over $g(x)$.

\subsection{Data Association and Mode-Model Association}
The difficulty of MMTT arises from both the unknown origin of measurements and the unknown motion mode of targets. Accordingly, two sorts of associations, data association and motion mode-model association, occur in MMTT.

Denote data association event $a_k^{i,j} = 1$ if measurement $j$ is originated from target $i$ at time $k$ and $a_k^{i,j} = 0$ otherwise.
In particular, if $j = 0$, $a_k^{i,j}$ represents the event that the detection of target $i$ is missed; if $i = 0$, $a_k^{i,j}$ represents the event that measurement $j$ is originated from clutter.
Note that $a_k^{0, 0}$ is meaningless so we let $p(a_k^{0,0} = 0) = 1$.
By the fact that data association events are mutually exclusive and exhaustive, a joint association event is defined as $A_k = \bigcap_{i = 0}^{N_T}\bigcap_{j = 0}^{N_{k, E}} a_{k}^{i, j}$.
In point target tracking, there often exists one-to-one frame (hard)~constraints in data association, that is, at each frame~(scan), a measurement can originate from at most one target or from clutter, and a target can generate at most one measurement. Based on the frame constraints, a joint event $A_k \in \mathcal{A}_k$ is \emph{feasible} if it fulfils the following equations, where $\mathcal{A}_k$ is the set of feasible joint association events.
\begin{equation}\label{frame}
\begin{split}
&\sum_{i = 0}^{N_T}a_k^{i,j} = 1, \quad \forall j \in \left \{1, \dots, N_{k,E} \right \}, \\
&\sum_{j = 0}^{N_{k,E}}a_k^{i,j} = 1, \quad \forall i \in \left \{1, \dots, N_{T} \right \}.
\end{split}
\end{equation}

Given $N_T$ targets and $N_{k,E}$ measurements, the prior probability of a joint association event $A_k$ is~\cite{NIPS2014}
\begin{align}\label{pirA}
p(A_k | E_k) = {(\lambda V_{G})^{N_{k,C}}\exp(-\lambda V_{G})}/{N_{k,E}!} \prod_{i=1}^{N_T} \left( P_d^i(e_k^i) \right)^{d_{k}^i} \left (1 - P_d^i(e_k^i) \right)^{1 - d_{k}^i},
\end{align}
where $d_{k}^i = 1 - a_k^{i,0}$ is the $i$th target detection indicator associated with $A_k$, $N_{k, C} = N_{k, E} - \sum_{i = 1}^{N_T}d_{k}^i$ is the number of clutter at time $k$.
In the vein of \cite{2009Blanding},
we use a two-value variable $P_d^i(e_k^i)$ to represent the time-varying and target visibility state-dependent detection probability.
Specifically, $P_d^i(e_k^i = 1) = p_d^i$ and $P_d^i(e_k^i = 0) = \varepsilon$, where $\varepsilon$ is a small positive real number (e.g., $\varepsilon = 0.1$). Note that the notion of $p_d^i$, which is a predefined value, represents the target detection probability in a single scan without considering the historic information on target visibility state.
%$P_d^i(e_k^i)$ is the target detection probability conditioned on target visibility state.
$P_d^i(e_k^i)$ equals $p_d^i$ if $p(e_k^i = 1) = 1$, in which case Eq.~(\ref{pirA}) is the same as the standard form~\cite{Bar1995Multitarget}.

Regarding motion mode-model association, the optimal approach to filter the state of the (hybrid)~system represented by Eqs.~(\ref{state}) and (\ref{measurement}) requires that every possible sequence of models from the beginning to the current time needs to be considered,
resulting in an exponentially increasing number of filters as the number of modes increases.

\subsection{Problem Statement}
Denote $X_{1:K}$, $E_{1:K}$, $M_{1:K}$, $A_{1:K}$ and $Y_{1:K}$ as batch sequences of target kinematic state, target visibility state, target motion model, data association and measurements from time $1$ to time $K$, respectively.
Let latent variables $\Theta_{1 : K} = \{X_{1:K}, E_{1:K}, M_{1:K}, A_{1:K}\}$.
The problem of MMTT is to estimate $X_{1:K}$ (tracking) and $E_{1:K}$ (detection) simultaneously, given $Y_{1:K}$ in the presence of unknown $A_{1:K}$ and $M_{1:K}$.

In the sense of Bayesian inference, the above joint detection and tracking problem is to calculate the joint posterior PDF $\mathcal{L}(\Theta_{1:K}) \triangleq p(\Theta_{1:K}|Y_{1:K})$ first, and then marginalize $\mathcal{L}(\Theta_{1:K})$ to obtain the posterior PDF of target kinematic state $X_{1:K}$ and posterior PDF of target visibility state $E_{1:K}$.
As~\cite{NIPS2014}, the interdependence among the latent variables is assumed as follows.
Target kinematic state $X_k$, target visibility state $E_k$ and motion mode-model association $M_k$ evolve with first-order Markov process.
Data association $A_k$ is independent over time.
At each time, measurement $Y_k$ is generated from $X_k$ via the measurement model,
and the relationship between target-to-measurement association is represented by $A_k$.
Additionally, $A_k$ is related to target visibility state $E_k$, and $X_k$ is conditionally independent of $E_k$ given $A_k$.  To this end, the full joint posterior PDF $\mathcal{L}(\Theta_{1:K})$ can be factorized into
\begin{equation}\label{L}
\begin{split}
\mathcal{L}(\Theta_{1:K}) \propto  \underbrace{\prod_{k = 1}^{K} \prod_{i = 1}^{N_T} \prod_{j = 1}^{N_{k,E}}V_G^{-a_k^{0, j}}\prod_{m_{k}^i = 1}^{N_M}  p(y_{k}^j|x_{k}^i, m_{k}^i, a_k^{i, j})^{a_k^{i,j}}}_{p_{Y_{1:K}|X_{1:K}, M_{1:K}, A_{1:K}}} \times \underbrace{\prod_{i = 1}^{N_T} p(x_{0}^i|m_0^i)\prod_{k = 1}^{K}  \prod_{m_{k}^i = 1}^{N_M}
p(x_{k}^i|x_{k-1}^i,m_{k}^i)}_{p_{X_{1:K}|M_{1:K}}} \\  \times \underbrace{\prod_{i = 1}^{N_T}\pi_{i,m} \prod_{k = 1}^K p(m_{k}^i|m_{k - 1}^i)}_{p_{M_{1:K}}} \times  \underbrace{\prod_{k = 1}^{K} p(A_k |E_k)}_{p_{A_{1:K}|E_{1:K}} }\mathbb{I}(A_k \in \mathcal{A}_k) \times \underbrace{\prod_{i = 1}^{N_T}\pi_{i,e} \prod_{k = 1}^K p(e_{k}^i|e_{k - 1}^i)}_{p_{E_{1:K}}}.
\end{split}
\end{equation}

If $A_{1:K}$ and $M_{1:K}$ are known, it is tractable to handle tracking by inferring the posterior PDF $p(X_{1:K} | M_{1:K}, A_{1:K}, Y_{1:K})$ via a fixed-interval smoother, and handle detection by inferring the posterior PDF $p(E_{1:K} |A_{1:K}, Y_{1:K})$ via a hidden Markov model~(HMM) smoother.
With unknown $A_{1:K}$ and $M_{1:K}$, however, it is required to summarize over all possible configurations of the data association hypotheses and motion mode-model hypotheses from $1$ to $K$.
Since the number of data association events increases exponentially with the number of targets and the number of (validated) measurements, and the number of motion mode-model hypotheses increases exponentially with the number of models, the exact calculation of joint posterior PDF is prohibitively expensive and approximation solutions are often sought.
Since the sampling-based stochastic approximation methods~(e.g., MCMC) are computationally intensive,
we adopt a combined BP-MF approximation in this paper.

%\textcolor{blue}{
%The sampling-based stochastic approximation methods~(e.g., MCMC) are computationally intensive. The analytical-based deterministic approximation methods, such as variational Bayes, are more computationally effective. Two kinds of typical variational Bayesian methods, i.e., MF approximation and BP, can be implemented in an iterative way, such as message passing. In the next section, we combine the MF approximation and BP into a unified message passing algorithm for simplifying the intractable inference problem arising from the multiple maneuvering target tracking.}

%Message-passing techniques, operating on factor graphs, are widely used to solve optimization and inference problem by iteratively exchanging information~(messages) with their neighbors~(connected nodes).
%Two kinds of analytical-based deterministic approximations including BP and MF approximation are popular to solving the intractable high-dimensional inference problem arising from the multiple maneuvering target tracking. Specifically, the high-dimensional joint posterior PDF $\mathcal{L}(\Theta_{1:K})$ is fully factorized by the product of individual PDFs of each latent variable via MF approximation, then the data association with one-to-one frame hard constraints is approximated by BP.  This splitting is convenient because BP works well with hard constraints and the MF approximation yields simple message passing update equations. It is of great benefit to applying BP and the MF approximation on the same factor graph to combine their respective virtues while circumventing their drawbacks.

\section{Solutions}\label{sec:solution}
\subsection{Combined BP-MF Approximation for MMTT}
MP techniques, operating on factor graphs, are broadly used to solve optimization and inference problem by iteratively exchanging information~(messages) between neighboring nodes.
BP is an explicit MP technique.
Many other inference algorithms, such as MF approximation, expectation-maximization, can also be interpreted as MP~\cite{winn2005varia, dauwels2005expe}.
MF approximates a joint distribution $p_{\bm{X}}(\bm{x})$ based on the minimization of the variational free energy,
which has the virtue of yielding closed-form computationally tractable expressions in conjugate-exponential models.
However, MF approximation fails if a factor graph has deterministic factor nodes, e.g., hard constraints.
BP computes the marginal distribution $p_i(x_i)$ of the variable $x_i$ associated to the joint distribution $p_{\bm{X}}(\bm{x})$ by minimizing the Bethe free energy, which works in models with deterministic factor nodes as well.
The fixed-point equations of both BP and MF approximation can be obtained via the region-based free energy approximation~\cite{yedidia2005constructing, riegler2012merging}.
Next, based on the region-based free energy approximation~\cite{yedidia2005constructing, riegler2012merging},
we describe the combined BP-MF approximation for MMTT.

\begin{figure}[!htp]
    \centering
    \includegraphics[width=1\textwidth]{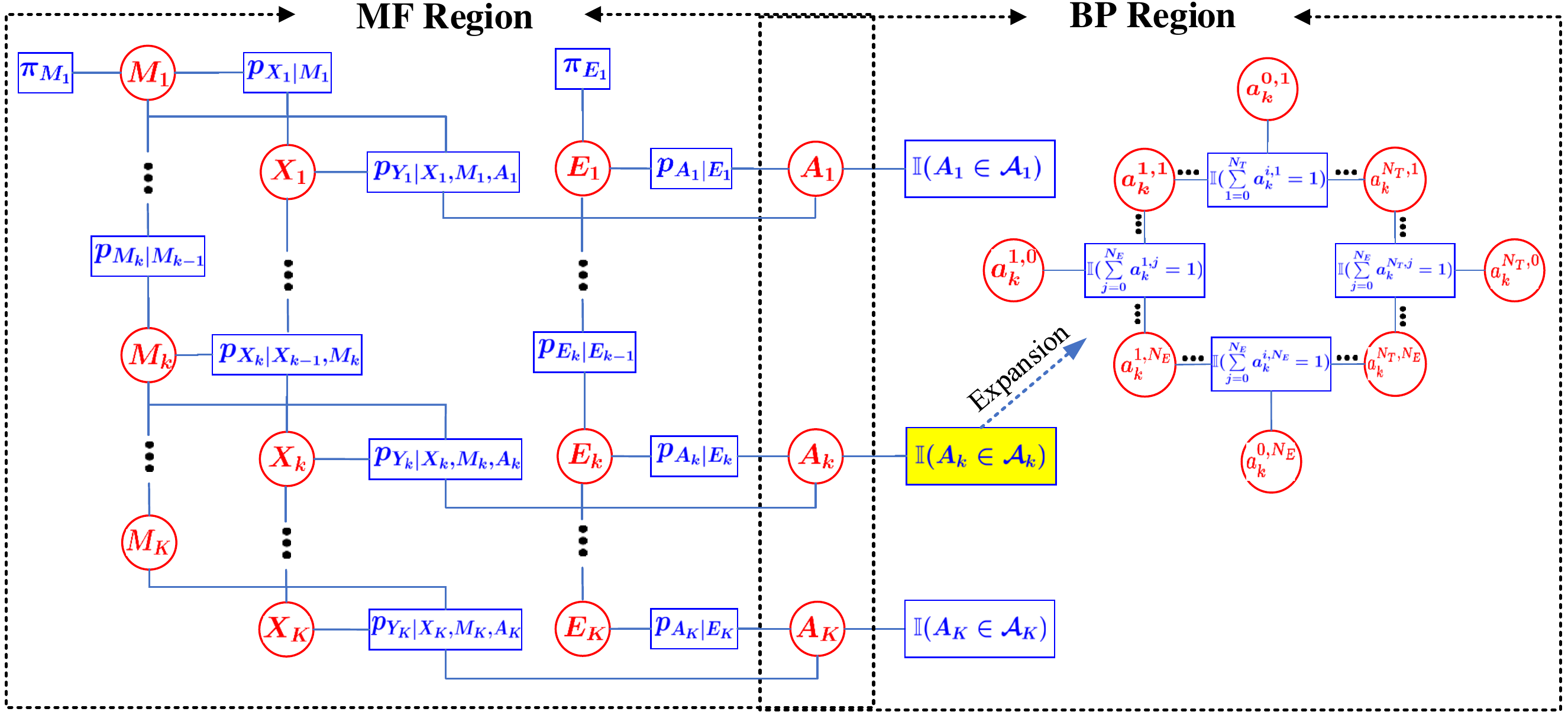}
    \caption{The factor graph corresponding to the factorization of the joint PDF in Eq.~(\ref{L}). The splitting of the factor graph into BP and MF part is chosen in such a way that utilizes most of the advantages of BP~(for hard constraints) and the MF approximation~(for conjugate-exponential model).}
    \label{Fig1}
\end{figure}

A factor graph~(e.g., Fig.~\ref{Fig1}) is a bipartite graph that has a variable node $i$~(typically represented by a circle) for each variable $x_i, i \in \mathcal{I}$, a factor node $a$~(represented by a square) for each factor $f_{a}, a \in \mathcal{F}$, with an edge connecting variable node $i$ to factor node $a$ if $x_i$ is an argument of $f_{a}$. In a factor graph, $\mathcal{S}(a)$ is the set of all variable nodes connected to a factor node $a \in \mathcal{F}$ and $\mathcal{S}(i)$ represents the set of all factor nodes connected to a variable node $i \in \mathcal{I}$.

Following the definitions in \cite{yedidia2005constructing, riegler2012merging}, a \emph{region} $R$ of a factor graph consists of subsets of indices $\mathcal{I}_R \subset \mathcal{I}$ and  $\mathcal{F}_R \subset \mathcal{F}$ with  the restriction that $a \in \mathcal{F}_R$ implies that $\mathcal{S}(a) \subseteq \mathcal{I}_R$.
Each region $R$ associates a \emph{counting number} $c_R \in \mathbb{Z}$.
For all $a\in \mathcal{F}$ and $i \in \mathcal{I}$, a set $\mathcal{R} = \{(R, c_R)\}$ of regions and associated counting numbers is called \emph{valid} if
\begin{equation}
  \sum_{(R, c_R) \in \mathcal{R}} c_R \mathbb{I}(a \in \mathcal{F}_R) = \sum_{(R, c_R) \in \mathcal{R}} c_R\mathbb{I}(i \in \mathcal{I}_R) = 1.
\end{equation}

By the fact that BP works well with hard constraints~(one-to-one frame constraints in data association) and the MF approximation yields simple MP update equations of conjugate-exponential model, the factor graph~(illustrated in Fig.~\ref{Fig1}) corresponding to MMTT can be divided into two regions, i.e., MF region $R_{\text{MF}} = (\mathcal{I}_{\text{MF}}, \mathcal{F}_{\text{MF}})$ and BP region $R_{\text{BP}} = (\mathcal{I}_{\text{BP}}, \mathcal{F}_{\text{BP}})$ with
\begin{align}\label{eq:splitFNodes}
\mathcal{I}_{\text{BP}} =& \{A_1, \ldots, A_K\}, \\
\mathcal{F}_{\text{BP}} =& \{\mathbb{I}(A_1 \in \mathcal{A}_1)\}\cup \cdots \cup \{\mathbb{I}(A_K \in \mathcal{A}_K)\}, \\
\mathcal{I}_{\text{MF}} =& \{X_1, \ldots, X_K\} \cup \{E_1, \ldots, E_K\} \cup \{M_1, \ldots, M_K\} \cup \{A_1, \ldots, A_K\}, \\
\mathcal{F}_{\text{MF}} =& \{p_{M_{1:K}}\} \cup \{p_{X_{1:K}|M_{1:K}} \} \cup \{p_{Y_{1:K}|X_{1:K}, M_{1:K}, A_{1:K}}\}  \cup \{p_{E_{1:K}}\} \cup \{p_{A_{1:K}|E_{1:K}}\}.
\end{align}
It is seen that $\mathcal{I}_{\text{BP}} \cup \mathcal{I}_{\text{MF}} = \mathcal{I}$, $\mathcal{I}_{\text{BP}} \cap \mathcal{I}_{\text{MF}} = \{A_1, \ldots, A_K\}$, $\mathcal{F}_{\text{BP}} \cup \mathcal{F}_{\text{MF}} = \mathcal{F}$ and $\mathcal{F}_{\text{BP}} \cap \mathcal{F}_{\text{MF}} = \emptyset$. The joint posterior PDF $\mathcal{L}(\Theta_{1:K})$ of Eq.(5) can be expressed as
\begin{equation}\label{RFG}
\mathcal{L}(\Theta_{1:K}) = \overbrace{p_{M_{1:K}} \times p_{X_{1:K}|M_{1:K}} \times p_{Y_{1:K}|X_{1:K}, M_{1:K}, A_{1:K}} \times
p_{E_{1:K}} \times p_{A_{1:K}|E_{1:K}}}^{\text{MF region}} \times \overbrace{\mathbb{I}(A_1 \in \mathcal{A}_1)\times \cdots \times \mathbb{I}(A_K \in \mathcal{A}_K)}^{\text{BP region}}.
\end{equation}

Let the counting number of MF region $c_{R_{\text{MF}}} = 1$.
The BP region is further divided into small regions $R_i = (\{i\}, \emptyset)$ with $c_{R_i} = 1 - |\mathcal{S}_{\text{BP}}(i)| - \mathbb{I}(i \in \mathcal{I}_{\text{MF}})$ for all $i \in \mathcal{I}_{\text{BP}}$, and large regions $R_a = (\mathcal{S}(a), \{a\})$ with $c_{R_a} = 1$ for all $a \in \mathcal{F}_{\text{BP}}$.
Accordingly, the joint posterior PDF $\mathcal{L}(\Theta_{1:K})$ can be approximated by minimizing the \emph{region-based free energy}~\cite{yedidia2005constructing}, (For continuous variables $x_i$, i.e., target kinematic state, one should replace the sum over $x_i$ by a Lebesgue integral.)
\begin{equation}
  F_{\text{BP, MF}} = \sum_{a \in \mathcal{F}_{\text{BP}}}\sum_{\bm{x}_a} b_a(\bm{x}_a) \ln \dfrac{b_a(\bm{x}_a)}{f_a(\bm{x}_a)} - \sum_{a \in \mathcal{F}_{\text{MF}}}\sum_{\bm{x}_a} \prod_{i \in \mathcal{S}(a)}b_i(x_i) \ln f_a(\bm{x}_a) - \sum_{i \in \mathcal{I}}(|\mathcal{S}_{\text{BP}}(i) - 1|) \sum_{x_i}b_i(x_i)\ln b_i(x_i),
\end{equation}
where the positive functions $b_a(\bm{x}_a)$ and $b_i(x_i)$, referred as beliefs, are the approximations of $f_a(\bm{x}_a)$ and $p(x_i)$, respectively. $b_a(\bm{x}_a)$ and $b_i(x_i)$ should fulfill the marginalization constraints and the normalization constraints as follows
\begin{equation}\label{eq:marconstraint}
b_i(x_i) = \sum_{\bm{x}_a \backslash x_i} b_a(\bm{x}_a), \quad \forall a \in \mathcal{F}_{\text{BP}}, i \in \mathcal{S}(a),
\end{equation}
and
\begin{equation}\label{eq:norconstraint}
\begin{split}
  &\sum_{x_i} b_i(x_i) = 1, \quad \forall i \in \mathcal{I}_{\text{MF}} \backslash \mathcal{I}_{\text{BP}}, \\
  &\sum_{\bm{x}_a} b_a(\bm{x}_a) = 1,  \quad \forall a \in \mathcal{F}_{\text{BP}}.
\end{split}
\end{equation}
Using the Lagrange multipliers method with the constraints given in Eqs.~(\ref{eq:marconstraint}), (\ref{eq:norconstraint}),
%and setting the derivatives of $F_{\text{BP, MF}}$ w.r.t. the beliefs $b_i(x_i)$ equal to zero,
a new MP scheme~\cite{riegler2012merging}, called BP-MF approach, is derived as follows.
\begin{equation}\label{Eq-update}
\begin{split}
n_{i \rightarrow a}(x_i) =& z_i \prod_{c \in \mathcal{S}_{\text{BP}}(i) \backslash a} m_{c \rightarrow i}^{\text{BP}}(x_i) \prod_{c \in \mathcal{S}_{\text{MF}}(i)} m_{c \rightarrow i}^{\text{MF}}(x_i), \quad \forall a \in \mathcal{F}, i \in \mathcal{S}(a) \\
 m_{a \rightarrow i}^{\text{BP}}(x_i) =& z_a \sum_{\bm{x}_a \backslash x_i} f_a(\bm{x}_a) \prod_{j \in \mathcal{S}(a) \backslash i} n_{j \rightarrow a}(x_j), \quad \quad \quad \;\; \forall a \in \mathcal{F_{\text{BP}}}, i \in \mathcal{S}(a) \\
 m_{a \rightarrow i}^{\text{MF}}(x_i) =& \exp\Big( \sum_{\bm{x}_a \backslash x_i} \prod_{j \in \mathcal{S}(a) \backslash i} n_{j \rightarrow a}(x_j) \ln  f_a(\bm{x}_a)  \Big), \;\; \forall a \in \mathcal{F_{\text{MF}}}, i \in \mathcal{S}(a)
\end{split}
\end{equation}
where $n_{i \rightarrow a}(x_i)$ is the message from the variable node $i$ to the factor node $a$, and $m_{a \rightarrow i}(x_i)$ is the message from the factor node $a$ to the variable node $i$. $z_i~(i \in \mathcal{I})$ and $z_a (a \in \mathcal{F}_{\text{BP}})$ are positive constants ensuring normalized beliefs. The notation $\mathcal{S}(a) \backslash i$ denotes the set of variable nodes that are neighbours of factor node $a$ but with variable node $i$ being removed, and $\sum_{\bm{x}_a \backslash x_i}$ denotes a sum over all the variables $\bm{x}_a$ that are arguments of $f_a$ except $x_i$.
%In fact, the variable-to-factor message $n_{i \rightarrow a}(x_i)~(a \in \mathcal{F}, i \in \mathcal{I})$ is an ``extrinsic'' value when $a \in \mathcal{F}_{\text{BP}}$ and a posterior probability when $a\in \mathcal{F}_{\text{MF}}$, i.e., $n_{i \rightarrow a}(x_i) = b_i(x_i)$ when $a \in \mathcal{F}_{\text{MF}}$.
Note that $n_{i \rightarrow a}(x_i) = b_i(x_i)$ when $a \in \mathcal{F}_{\text{MF}}$.

The belief $b_i(x_i)$ at a variable node $i$, which is the approximation to the exact marginal probability function $p_i(x_i)$, can be computed from the equation
\begin{equation}\label{eq:bx}
b_i(x_i) = z_i \prod_{a \in \mathcal{S}_{\text{BP}}(i)} m_{a \rightarrow i}^{\text{BP}}(x_i) \prod_{a \in \mathcal{S}_{\text{MF}}(i)} m_{a \rightarrow i}^{\text{MF}}(x_i), \quad \forall i \in \mathcal{I}.
\end{equation}

In the remainder of this section, we will present the detailed derivations of each beliefs together with
the corresponding subgraphs of the factor graph in Fig.~\ref{Fig1} to show the related variable nodes, factor nodes and messages more clearly.

\subsubsection{Derivation of Belief~$b_X(X)$}
\begin{figure}[!htp]
\centering
\includegraphics[width=0.54\textwidth]{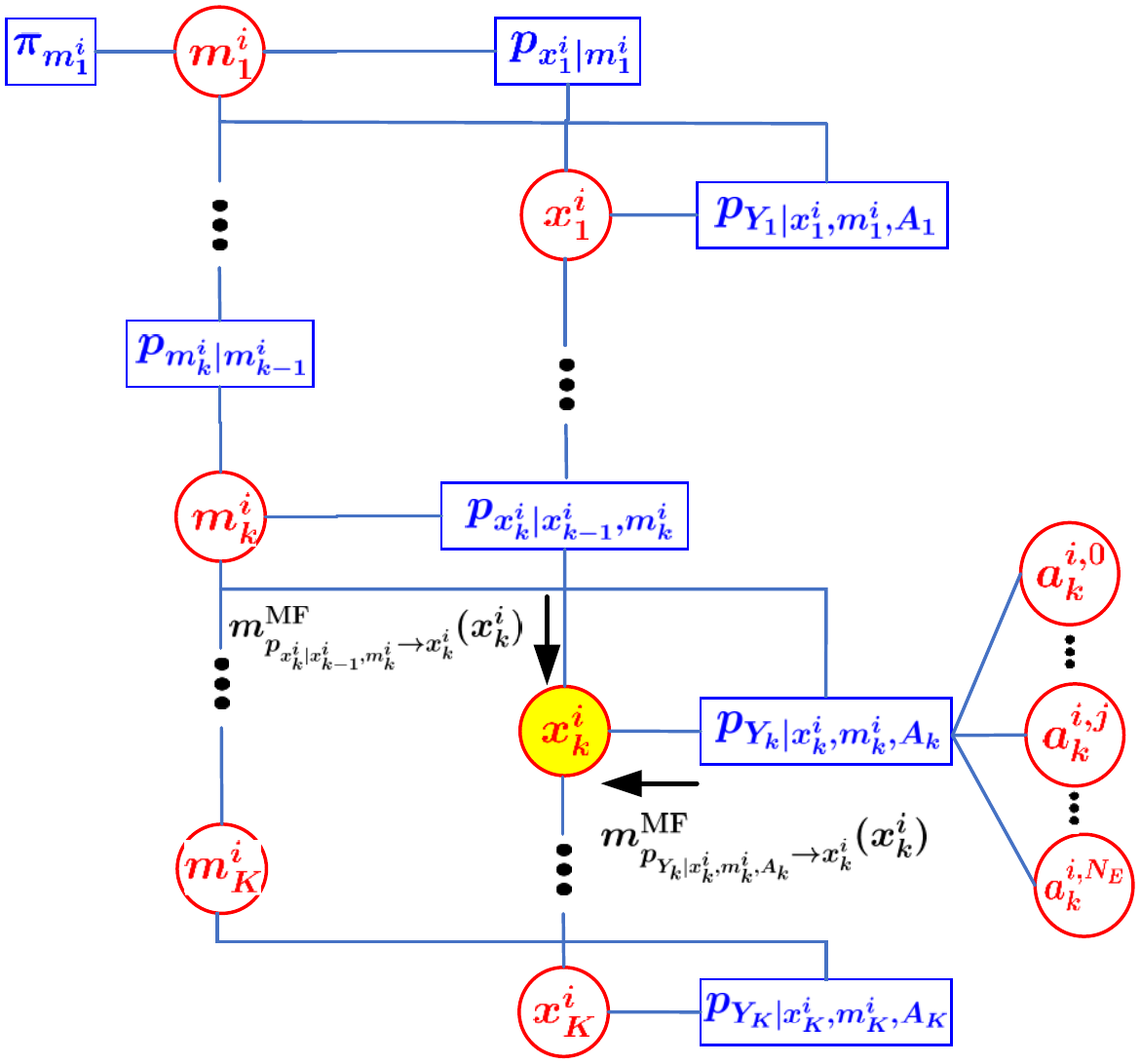}
\caption{The target kinematic state estimation subgraph.}
\label{Fig2}
\end{figure}

Based on the assumption that each target moves independently, the belief of kinematic state of all targets can be factorized as,
\begin{equation}\label{eq:bKS}
b_X(X) = \prod_{i = 1}^{N_T}b_X(x^i_{1:K}) = \prod_{i = 1}^{N_T} \prod_{k=1}^K b_X(x^i_k).
\end{equation}
%The target kinematic state estimation subgraph of target $i$ at time $k$ is the graphical representation of the belief $b_X(x^i_k)$.
Fig.~\ref{Fig2} shows the target kinematic state estimation subgraph that corresponding to the belief $b_X(x^i_k)$.

In Fig.~\ref{Fig2}, $x_k^i, i = 1, \ldots, N_T, k = 1, \ldots, K$, are the variable nodes to be considered.
Our aim is to calculate belief $b_X(x_k^i)$.
For each variable node $x_k^i$, connect it with two factor nodes,
$\mathcal{S}(x_i^k) = \big\{p_{x_k^i|x_{k-1}^i,m_{k}^i}, p_{y_k|x_k^i, m_k^i, A_k}\big\}$.
The sets of variable nodes connected to the each factor node are $\mathcal{S}(p_{x_k^i|x_{k-1}^i,m_{k}^i}) = \{x_k^i, x_{k-1}^i, m_{k}^i\}$ and $\mathcal{S}(p_{y_k|x_k^i, m_k^i, A_k}) = \{x_k^i, m_{k}^i, A_k\}$.
According to Eq.~(\ref{eq:bx}), the belief $b_X(x_k^i)$ can be computed by multiplying all the incoming messages
from the factor nodes $\mathcal{S}(x_i^k)$ to the variable node $x_k^i$, that is,
\begin{align}\label{eq:bKSx}
b_X(x_k^i) \propto m^{\text{MF}}_{p_{x_k^i|x_{k-1}^i,m_{k}^i} \rightarrow x_k^i}(x_k^i) \times m^{\text{MF}}_{p_{y_k|x_k^i, m_k^i, A_k} \rightarrow x_k^i}(x_k^i).
\end{align}
Using the message-computation rules given in Eq.~(\ref{Eq-update}), the factor-to-variable messages in Eq.~(\ref{eq:bKSx}) are calculated as
\begin{align}
m^{\text{MF}}_{p_{x_k^i|x_{k-1}^i,m_{k}^i} \rightarrow x_k^i}(x_k^i) =& \exp\left(\int_{x_{k-1}^i} \sum_{m_k^i = 1}^{N_M} n_{x_{k-1}^i \rightarrow p_{x_k^i|x_{k-1}^i, m_k^i}}(x_{k-1}^i) n_{m_{k}^i \rightarrow p_{x_k^i|x_{k-1}^i, m_k^i}}(m_{k}^i) \ln \mathcal{N}\left(x_k^i| F^{i,m}_k x_{k-1}^i, Q_k^{i,m}\right) d_{x_{k-1}^i} \right) \label{eq:F2VM}, \\
m^{\text{MF}}_{p_{y_k|x_k^i, m_k^i, A_k} \rightarrow x_k^i}(x_k^i) =& \exp \left(\sum_{m_k^i = 1}^{N_M}\sum_{j = 1}^{N_{k,E}} n_{m_{k}^i \rightarrow p_{y_k|x_k^i, m_k^i, A_k}}(m_{k}^i) n_{a_{k}^{i,j} \rightarrow p_{y_k|x_k^i, m_k^i, A_k}}(a_{k}^{i,j}) \ln \mathcal{N}\left(y_k^j| h_k^m(x_k^i), R_k^{i,m}\right)^{a_k^{i,j}}\right) \label{eq:F2VM2}.
\end{align}
Recall that for all $a \in \mathcal{F}_{\text{MF}}$ and $i \in \mathcal{S}(a)$, the variable-to-factor messages $n_{i \rightarrow a}(x_i) = b_i(x_i)$~\cite{riegler2012merging}. Thus, we have
\begin{align}
n_{x_{k-1}^i \rightarrow p_{x_k^i|x_{k-1}^i, m_k^i}}(x_{k-1}^i) =& b_X(x_{k-1}^i), \quad
n_{m_{k}^i \rightarrow p_{x_k^i|x_{k-1}^i, m_k^i}}(m_{k}^i)  = b_M(m_k^i) \label{eq:V2FM}, \\
n_{m_{k}^i \rightarrow p_{y_k|x_{k}^i, m_k^i, A_k}}(m_{k}^i) =& b_M(m_k^i), \quad
n_{a_{k}^{i,j} \rightarrow p_{y_k|x_k^i, m_k^i, A_k}}(a_{k}^{i,j}) = b_A(a^{i,j}_k) \label{eq:V2FM2}.
\end{align}

Substituting Eqs.~(\ref{eq:V2FM}), (\ref{eq:V2FM2}) into Eqs.~(\ref{eq:F2VM}), (\ref{eq:F2VM2}), we have
\begin{align}
m^{\text{MF}}_{p_{x_k^i|x_{k-1}^i,m_{k}^i} \rightarrow x_k^i}(x_k^i) \propto& \prod_{m_k^i = 1}^{N_M} \mathcal{N}\left(x_k^i|F_k^{i,m}\hat x_{k-1}^i, F_k^{i, m} P^i_{k-1}(F_k^{i, m})^T + Q_k^{i,m}\right)^{\hat m_k^i} \label{eq:mMFpx2x}, \\
m^{\text{MF}}_{p_{y_k|x_k^i, m_k^i, A_k} \rightarrow x_k^i}(x_k^i) \propto& \prod_{m_k^i = 1}^{N_M} \prod_{j = 1}^{N_{k,E}} \mathcal{N}\left(y_k^j| h_k^m(x_k^i), R_k^{i,m}\right)^{\hat m_k^i, \hat a_k^{i,j}}
                                                            \propto \prod_{m_k^i = 1}^{N_M} \mathcal{N} \left(\bar y_k^i| h_k^m(x_k^i), \bar R_k^{i,m}\right)^{\hat m_k^i} \label{eq:mMFpx2x2},
\end{align}
where $\hat m_k^i = \langle m^i_k \rangle_{b_M(m_k^i)}$, $\hat x_{k-1}^i =\langle x_{k-1}^i \rangle_{b_X(x_k^i)}$ and $\hat a_k^{i,j} = \langle a^{i,j}_k \rangle_{b_A(a_k^{i,j})}$ are the expectations of $m_k^i$, $x_{k-1}^i$ and $a_k^{i,j}$ taken over corresponding beliefs.
The synthetic measurement $\bar y_k^i$ and the corresponding covariance matrix $\bar R_k^{i, m}$ in Eq.~(\ref{eq:mMFpx2x2}) are defined as
\begin{equation} \label{App-bar-y}
\bar{y}_{k}^i =  \dfrac{\sum \nolimits_{j=1}^{N_{k,E}} \hat a_{k}^{i,j}y_{k}^j}{1- \hat a_{k}^{i,0}},\quad
\bar{R}_{k}^{i, m} = \frac{R_{k}^{i,m}}{1- \hat a_{k}^{i,0}}.
\end{equation}

Substituting Eqs.~(\ref{eq:mMFpx2x}), (\ref{eq:mMFpx2x2}) into Eq.~(\ref{eq:bKSx}), the belief $b_X(x_k^i)$ is rewritten as
\begin{align}\label{eq:bKSxUpdate}
b_X(x_k^i) \propto& \prod_{m_k^i = 1}^{N_M} \mathcal{N}\left(x_k^i|F_k^{i,m}\hat x_{k-1}^i, F_k^{i, m} P^i_{k-1}(F_k^{i, m})^T + Q_k^{i,m}\right)^{\hat m_k^i} \times \prod_{m_k^i = 1}^{N_M}\mathcal{N}\left(\bar y_k^i| h_k^m(x_k^i), \bar R_k^{i,m}\right)^{\hat m_k^i} \\ \nonumber
           \propto& \prod_{m_k^i = 1}^{N_M} \mathcal{N}\left(x_k^{i}|\hat x_{k}^{i, m}, P_{k}^{i, m} / \hat m_k^i  \right).
\end{align}

Eq.~(\ref{eq:bKSxUpdate}) shows that, for each target $i$, posterior PDF~(belief) $b_X(x_k^i)$ is Gaussian distributed with the product form of mode-dependent PDFs $b_X(x^{i,m}_k)$, $m = 1,2,\ldots, N_M$. The maximum a posteriori estimation of the parameters of $b_M(x^{i,m}_{k})$ can be achieved by a Kalman filter~(for linear models) or a nonlinear filter~(for nonlinear models) on an averaged state space system with synthetic measurement $\bar y_k^i$ and covariance $\bar R_k^{i,m}$,
that is,
\begin{equation}\label{eq:xP}
\hat x_k^{i, m} = \mathbb{E}\left({x}_{k}^{i, m}|\bar{y}_{k}^i, m_{k}^i\right), \quad P_k^{i, m} = \text{cov}\left(\hat{x}_{k}^{i, m},\hat{x}_{k}^{i, m}|\bar{y}_{k}^i, \bar R_{k}^{i, m}, m_{k}^i\right).
\end{equation}

The posterior PDF~(belief) $b_X(x_k^i)$ is then obtained by fusing the local mode-dependent PDFs $b_X(x^{i,m}_k)$, $m = 1,2,\ldots, N_M$, with mean $\hat x_{k|1:K}^i$ and covariance $\hat P_{k|1:K}^i$ given by
\begin{equation}\label{E-x}
\hat x_k^i = P_k^i \sum_{m^i_k = 1}^{N_M}\hat m_k^i \left( P_k^{i, m}\right)^{-1} \hat x_k^{i, m}, \quad \left( P_k^i \right)^{-1} = \sum_{m^i_k = 1}^{N_M} \hat m_k^i \left( P_k^{i, m}\right)^{-1}.
\end{equation}

For a time sequence $1:K$, the belief $b_X(x_{1:K}^i)$ is derived as
\begin{equation}\label{eq:bx1k}
b_X(x_{1:K}^i) =  \prod_{k=1}^K b_X(x^i_k) = \prod_{k=1}^K \mathcal{N}\left(x_k^{i}|\hat x_{k}^{i, m}, P_{k}^{i, m} / \hat m_k^i  \right).
\end{equation}
In this case, one simply replaces the filtering in Eq.~(\ref{eq:xP}) with smoothing,
and the Kalman smoother~(for linear models) or nonlinear smoother such as Unscented Rauch-Tung-Striebel Smoother~(URTS)\cite{sarkka2013bayesian} can be used.

\subsubsection{Derivation of Belief $b_E(e)$}
Like the target kinematic state, the belief of the visibility state of all targets can be factorized as,
\begin{equation}\label{eq:bVS}
b_E(E) = \prod_{i = 1}^{N_T}b_E(e^i_{1:K}) = \prod_{i = 1}^{N_T} \prod_{k=1}^K b_E(e^i_k).
\end{equation}

\begin{figure}[!htp]
\centering
\includegraphics[width=0.25\textwidth]{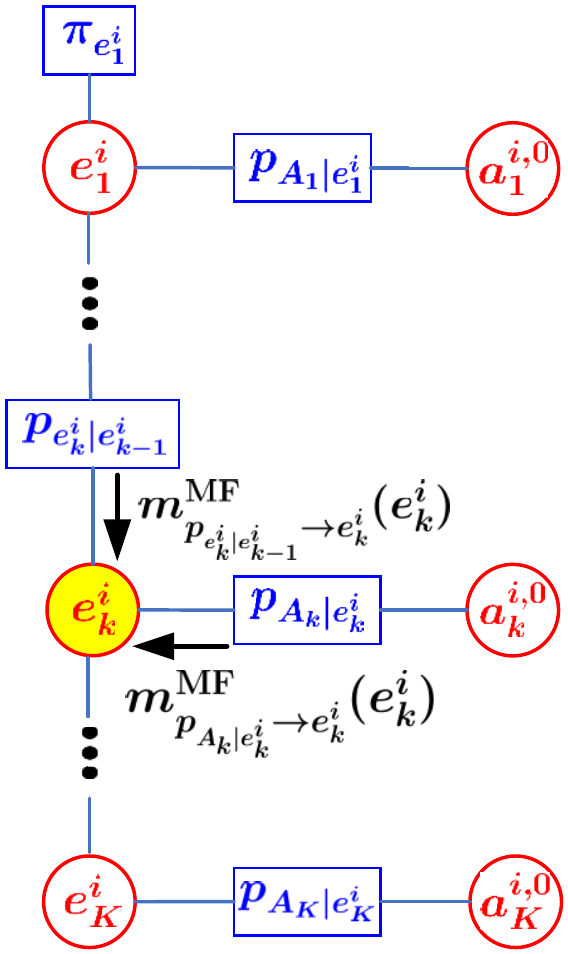}
\caption{The target visibility state estimation subgraph.}
\label{Fig3}
\end{figure}

Fig.~\ref{Fig3} shows the target visibility state estimation subgraph corresponding to the belief $b_E(e_k^i)$.
The to-be-considered variable nodes of the target detection subgraph are $e_k^i, i = 1, \ldots, N_T, k = 1, \ldots, K$.
For each variable node $e_k^i$, connect it with two factor nodes,
$\mathcal{S}(e_k^i) = \big\{p_{e_k^i|e_{k-1}^i}, p_{A_k|e_k^i}\big\}$.
The sets of variable nodes connected to each factor node are $\mathcal{S}(p_{e_k^i|e_{k-1}^i}) = \{e_k^i, e_{k-1}^i\}$ and $\mathcal{S}(p_{A_k|e_k^i}) = \{e_k^i, A_k\}$, respectively.
According to the message-computation rules given in Eq.~(\ref{Eq-update}),
the messages from the factor nodes $\mathcal{S}(e_k^i)$ to the variable node $e_k^i$ are calculated as
\begin{align}
m^{\text{MF}}_{p_{e_k^i|e_{k-1}^i} \rightarrow e_k^i}(e_k^i) = & \exp\Big(\sum_{e_{k-1}^i = 0}^1 n_{e_{k-1}^i \rightarrow p_{e_k^i|e_{k-1}^i}}(e_{k-1}^i) \ln T_{i,e}(e_{k-1}^i, e_{k}^i) \Big) = T_{i,e}(e_{k-1}^i, e_{k}^i), \\
m^{\text{MF}}_{p_{A_k|e_k^i} \rightarrow e_k^i}(e_k^i) = & \exp \Big(\sum_{a_k^{i,0} = 0}^1 n_{a_k^{i,0} \rightarrow p_{A_k|e_k^i}}(a_k^{i,0}) \ln p(A_k|e_k^i) \Big)
                                                \propto \exp \left((1 - \hat a_k^{i,0}) \ln(P_d^i(e_k^i)) + \hat a_k^{i,0} \ln(1 - P_d^i(e_k^i)) \right).
\end{align}
According to Eq.~(\ref{eq:bx}), the belief $b_E(e_k^i)$ can be computed as
\begin{align}
b_E(e_k^i) \propto & m^{\text{MF}}_{p_{e_k^i|e_{k-1}^i} \rightarrow e_k^i}(e_k^i) \times m^{\text{MF}}_{p_{A_k|e_k^i} \rightarrow e_k^i}(e_k^i) \\ \nonumber
            = & T_{i,e}(e_{k-1}^i, e_{k}^i) \underbrace{\exp \left((1 - \hat a_k^{i,0}) \ln(P_d^i(e_k^i)) + \hat a_k^{i,0} \ln(1 - P_d^i(e_k^i)) \right)}_{\xi_{k}(e_k^i)}.
\end{align}

For a time sequence $1:K$, the belief $b_E(e_{1:K}^i)$ is derived as
\begin{equation}\label{eq:bee1k}
b_E(e_{1:K}^i) = \prod_{k = 1}^K b_E(e_k^i) = \pi_{e_1^i} \xi_{1}(e_1^i) \prod_{k = 2}^K T_{i,e}(e_{k-1}^i, e_{k}^i) \xi_{k}(e_k^i).
\end{equation}
It is seen that the belief $b_E(e_{1:K}^i)$ follows a HMM with the indirect observation sequence $\{\xi_{1}(e_1^i), \ldots, \xi_{K}(e_K^i)\}$, and the estimation of $b_E(e_{1:K}^i)$ can be addressed by a forward-backward algorithm~\cite{rabiner1989tutorial}.

\subsubsection{Derivation of Belief $b_M(m)$}
\begin{figure}[!htp]
\centering
\includegraphics[width=0.54\textwidth]{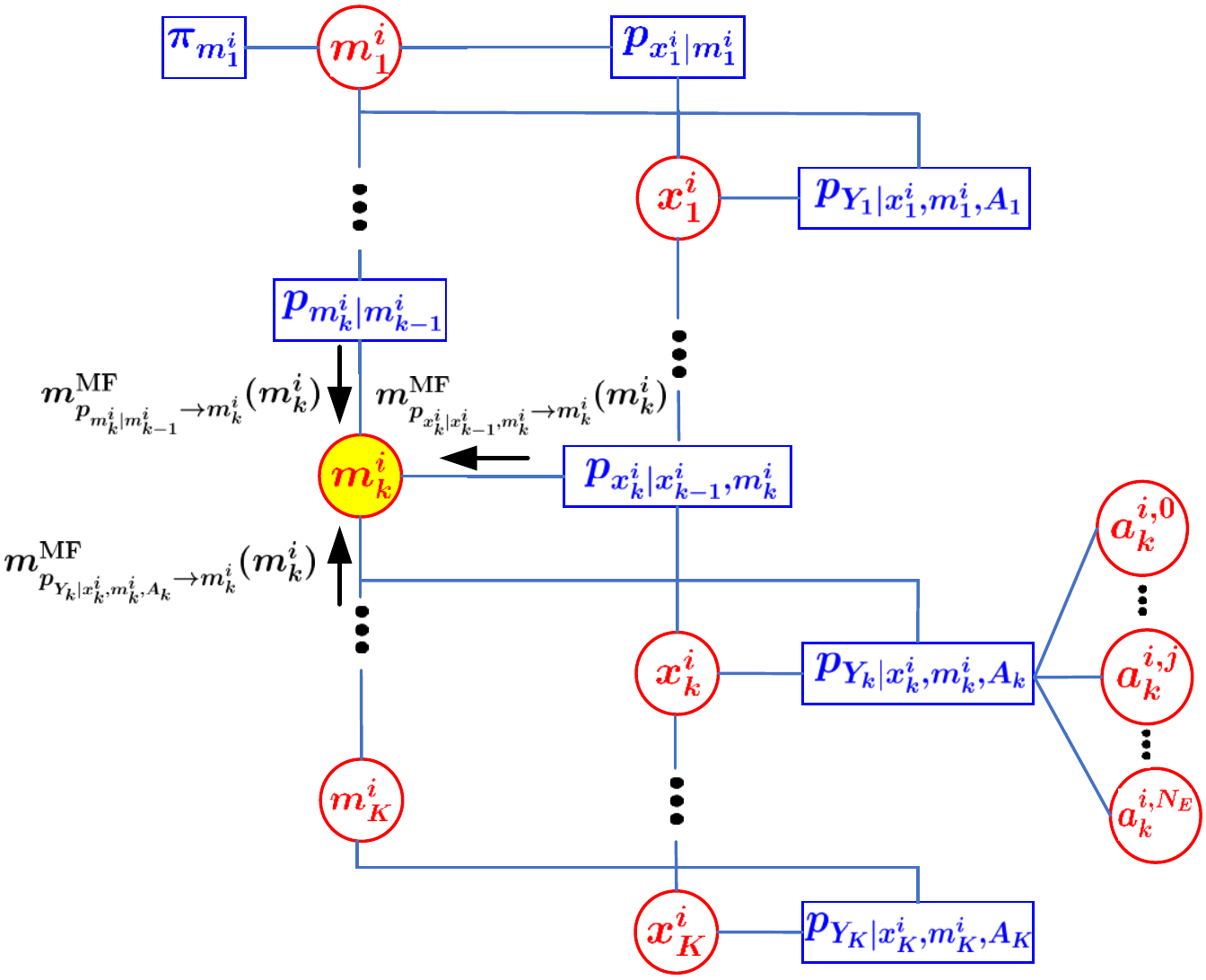}
\caption{The target motion mode-model association subgraph.}
\label{Fig4}
\end{figure}

Similarly, the belief $b_M(M)$ is factorized over targets.
The subgraph of the target motion mode-model association corresponding to the belief $b_M(m_k^i)$ is illustrated in Fig.~\ref{Fig4}.
In Fig.~\ref{Fig4}, $m_k^i, i = 1, \ldots, N_T, k = 1, \ldots, K$, are to-be-considered variable nodes.
For each $m_k^i$, connect it with three factor nodes, $\mathcal{S}(m_k^i) = \{p_{m_k^i|m_{k-1}^i}, p_{x_k^i|x_{k-1}^i, m_{k}^i}, p_{y_k|x_k^i, m_k^i, A_k^i}\}$.  The sets of variable nodes connected to each factor node are $\mathcal{S}(p_{m_k^i|m_{k-1}^i}) = \{m_k^i, m_{k-1}^i\}$, $\mathcal{S}(p_{x_k^i|x_{k-1}^i, m_{k}^i}) = \{x_k^i, x_{k-1}^i, m_{k}^i\}$, and $\mathcal{S}(p_{y_k|x_k^i, m_k^i, A_k}) = \{x_k^i, m_{k}^i, A_k\}$.
The messages from each factor nodes $\mathcal{S}(m_k^i)$ to variable node $m_k^i$ is calculated as
\begin{align}
m^{\text{MF}}_{p_{m_k^i|m_{k-1}^i} \rightarrow m_k^i}(m_k^i) =& \exp\Big(\sum_{m_{k-1}^i = 1}^{N_M} n_{m_{k-1}^i \rightarrow p_{m_k^i|m_{k-1}^i}}(m_{k-1}^i) \ln T_{i,m}(m_{k-1}^i, m_{k}^i) \Big)
                                                      = T_{i,m}(m_{k-1}^i, m_{k}^i), \\
m^{\text{MF}}_{p_{x_k^i|x_{k-1}^i, m_{k}^i} \rightarrow m_k^i}(m_k^i) =& \exp \Bigg( \underbrace{\int_{x_{k}^i}\int_{x_{k-1}^i}n_{x_{k}^i \rightarrow  p_{x_k^i|x_{k-1}^i, m_k^i}}(x_{k}^i)n_{x_{k-1}^i \rightarrow p_{x_k^i|x_{k-1}^i, m_k^i}}(x_{k-1}^i) \ln \mathcal{N}\left(x_k^i|F_k^{i,m}x_{k-1}^i, Q_k^{i,m}\right) d_{x_k^i}d_{x_{k-1}^i}}_{\mathcal{M}_{x,k}}
\Bigg), \\
m^{\text{MF}}_{p_{y_k|x_k^i, m_k^i, A_k^i} \rightarrow m_k^i}(m_k^i) =& \exp \Bigg( \underbrace{\int_{x_{k}^i}\sum_{j = 1}^{N_{k,E}}n_{x_{k}^i \rightarrow p_{y_k|x_k^i, m_k^i, A_k^i}}(x_{k}^i)n_{a_{k}^{i,j} \rightarrow p_{y_k|x_k^i, m_k^i, A_k^i}}(a_k^{i,j}) \ln \mathcal{N}\left(y_k^j|h_k^{m}(x_k^i), R_k^{i,m}\right)^{a_k^{i,j}} d_{x_k^i}}_{\mathcal{M}_{y,k}}
\Bigg)
\end{align}
with
\begin{align}\label{eq:mxk}
\mathcal{M}_{x,k} \triangleq& -\dfrac{1}{2} \text{Tr}\left\{\left(P_{k|k-1}^{i,m}\right)^{-1} \left\langle\left(x_k^i - F_k^{i,m}x_{k-1}^i\right)\left(x_k^i - F_k^{i,m}x_{k-1}^i\right)^T\right\rangle_{b_X(x_{k}^i),b_X(x_{k-1}^i)} \right\} \\ \nonumber
=&  -\dfrac{1}{2} \text{Tr}\left\{\left(P_{k|k-1}^{i,m}\right)^{-1} \left(P_k^i - P_{k,k-1}(F_k^{i,m})^T - F_k^{i,m}P_{k,k-1} + F_k^{i,m}P_{k-1}^i(F_k^{i,m})^T + (\hat x_k^i - F_k^{i,m}\hat x_{k-1}^i) (\hat x_k^i - F_k^{i,m}\hat x_{k-1}^i)^T \right) \right\} \\
\label{eq:myk} \mathcal{M}_{y,k} \triangleq& -\dfrac{1}{2} \text{Tr}\Big\{\left(S_{k}^{i,m}\right)^{-1}  \hat a_k^{i,j} \sum_{j = 1}^{N_{k,E}} \left\langle \left(y_k - h_k^m(x_k^i)\right)\left(y_k - h_k^m(x_k^i)^T\right)\right\rangle_{b_X(x_{k}^i)} \Big\} \\ \nonumber
=& -\dfrac{1}{2} \text{Tr}\left\{\left(S_{k}^{i,m}\right)^{-1} \left( (\bar y_k^i - h_k^m(\hat x_k^i))(\bar y_k^i - h_k^m(\hat x_k^i))^T + H_k^mP_k^i(H_k^m)^T \right) \right\}
\end{align}
In Eqs.~(\ref{eq:mxk})-(\ref{eq:myk}), $P_{k|k-1}^{i,m}$ is the predicted covariance of $x_{k}^{i,m}$, $P^i_{k,k-1}$ is the covariance of $x_{k}^i$ and $x_{k-1}^i$, $S_k^{i,m}$ is the innovation covariance, $\hat x_k^i = \langle x_k^i \rangle_{b_X(x_k^i)}$ is the expectation of $x_k^i$ taken over the beliefs $b_X(x_k^i)$, and $H_k^m$ is the Jacobian matrix of function $h_k^m$ with respect to~(w.r.t.) $x$.

According to Eq.~(\ref{eq:bx}), the belief $b_M(m_k^i)$ can be computed as
\begin{align}
b_M(m_k^i) \propto & m^{\text{MF}}_{p_{m_k^i|m_{k-1}^i} \rightarrow m_k^i}(m_k^i) \times m^{\text{MF}}_{p_{x_k^i|x_{k-1}^i, m_{k}^i} \rightarrow m_k^i}(m_k^i) \times m^{\text{MF}}_{p_{y_k|x_k^i, m_k^i, A_k^i} \rightarrow m_k^i}(m_k^i) \\ \nonumber
= & T_{i,m}(m_{k-1}^i, m_{k}^i) \underbrace{\exp\left(\mathcal{M}_{x,k} + \mathcal{M}_{y,k}\right)}_{\mathcal{M}_{k}(m_k^i)}.
\end{align}

The belief $b_M(m_{1:K}^i)$ over a time sequence $1:K$ is thus given by
\begin{equation}\label{eq:bMm}
b_M(m_{1:K}^i)  = \pi_{m_1^i} \mathcal{M}_{1}(m_1^i) \prod_{k = 2}^K T_{i,m}(m_{k-1}^i, m_{k}^i) \mathcal{M}_{k}(m_k^i).
\end{equation}
From Eq.~(\ref{eq:bMm}), it is seen that the belief $b_M(m_{1:K}^i)$ follows a HMM with observation sequence $\{\mathcal{M}_{1}(m_1^i), \ldots, \mathcal{M}_{K}(m_K^i)\}$, and the estimation of $b_M(m_{1:K}^i)$  can be addressed by a forward-backward algorithm as well~\cite{rabiner1989tutorial}.

\subsubsection{Derivation of Belief $b_A(a)$}
In this paper, we assume that the data association is independent over different scans.
Accordingly, the belief on data association $b_A(a^{i,j}_k)$ is factorized over time horizon, that is,
\begin{equation}
b_A(A) = \prod_{k=1}^K b_A(a_k).
\end{equation}

\begin{figure}[!htp]
\centering
\includegraphics[width=0.48\textwidth]{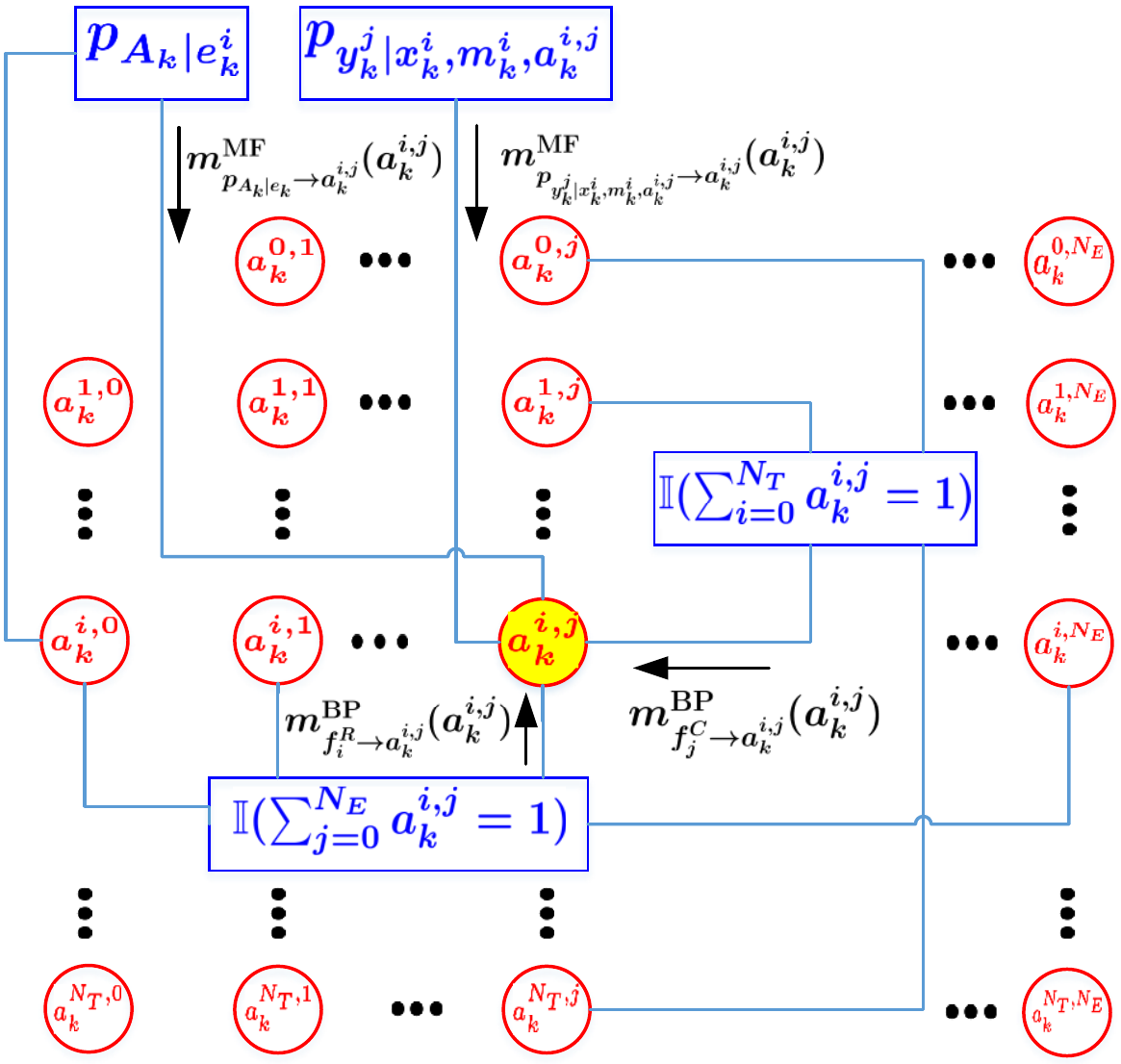}
\caption{The data association subgraph.}
\label{Fig5}
\end{figure}

Fig.~\ref{Fig5} shows the data association subgraph corresponding to the belief $b_A(a)$.
The data association subgraph consists of the variable nodes $a_k^{i,j}, i = 0, 1, \ldots, N_T, j = 0, \ldots, N_{k,E}, k = 1, \ldots, K$.
There are four factor nodes neighbouring to variable node $a^{i,j}_k$, i.e., $\mathcal{S}(a^{i,j}_k) = \big \{p_{A_k|e_k}, p_{y^j_k|x_k, m_k, a^{i,j}_k}, f_i^R, f_j^C \big \}$ where we denote $f_i^R = \mathbb{I}\big(\sum_{i = 0}^{N_T} a_k^{i,j} = 1\big)$ and $f_j^C = \mathbb{I}\big(\sum_{j = 0}^{N_{k,E}} a_k^{i,j} = 1\big)$ for simplicity. The sets of variable nodes connected to the corresponding factor node are $\mathcal{S}(p_{A_k|e_k}) =  \{A_k, e_k  \}$, $\mathcal{S}(p_{y^j_k|x_k, m_k, a^{i,j}_k}) =  \{y_k^j, x_{k}^i, a_{k}^{i,j}  \}$, $\mathcal{S}(f_i^R) =  \{a_k^{0,j}, \ldots, a_k^{N_T, j}  \}$ and $\mathcal{S}(f_j^C) =  \{a_k^{i, 0}, \ldots, a_k^{i, N_{k,E}}  \}$.

By the message-computation rules given in Eq.~(\ref{Eq-update}), the messages from each factor nodes $\mathcal{S}(a^{i,j}_k)$ to the variable node $m_k^i$ is calculated as follows.

For the messages belong to the MF region ($a \in \mathcal{F}_{\text{MF}}$), we have
\begin{equation}
\begin{split}
m^{\text{MF}}_{p_{A_k|e_k} \rightarrow a^{i,j}_k}(a^{i,j}_k) =& \exp\left(\sum_{e_k^i = 0}^1 n_{e_k^{i} \rightarrow p_{A_k|e_k}} (e_k^i) \ln p(A_k|e_k^i) \right)\\
                                          \propto & \exp \left( \left\langle  \sum_{j = 1}^{N_{k,E}} a_k^{i,j} \ln P_d^i(e_k^i) + a_k^{i,0} \ln (1 - P_d^i(e_k^i)) \right\rangle_{b_E(e_k^i)} \right) \\
                                          =& \begin{cases}
                                          \left\langle 1 - P_d^i(e_k^i) \right\rangle_{b_E(e_k^i)}^{a_k^{i,j}}, & \forall i > 0, j = 0 \\
                                          \left\langle P_d^i(e_k^i) \right\rangle_{b_E(e_k^i)}^{a_k^{i,j}}, & \forall i > 0, j >0
                                          \end{cases}
\end{split}
\end{equation}
\begin{equation}
\begin{split}
m^{\text{MF}}_{p_{y^j_k|x_k, m_k, a^{i,j}_k} \rightarrow a^{i,j}_k}(a^{i,j}_k) =& \exp \left(\int_{x_k^i} \sum_{m_k^i = 1}^{N_M} n_{x_k^i \rightarrow p_{y^j_k|x_k, m_k, a^{i,j}_k}}(x_k^i)n_{m_k^i \rightarrow p_{y^j_k|x_k, m_k, a^{i,j}_k}}(m_k^i) \ln p(y_k^j|x_k^i, a_k^{i,j}, m_k^i) d_{x_k^i} \right)\\
                                                    \propto & \exp \Bigg( a_k^{0, j} \ln (1 / V_G) + a_k^{i,j} \underbrace{\left\langle \ln \mathcal{N}\left(y_k^j|h_k^m(x_k^i), S_k^{i,m}\right) \right\rangle_{x_k^i, m_k^i}}_{\mathcal{X}_k} \Bigg) \\
                                                    =& \begin{cases}
                                                    {V_G}^{-a_k^{0,j}}, & \forall i = 0, j > 0 \\
                                                    \exp(a_k^{i,j}\mathcal{X}_k),  & \forall i >0, j>0
                                                    \end{cases}
\end{split}
\end{equation}
with
\begin{equation}
\mathcal{X}_k =- \biggl\{\frac{1}{2} \text{Tr} \Bigl\{(S_{k}^{i, m})^{-1}\bigl((y_{k}^j - h^m_k(\hat{x}_{k}^i))(y_{k}^j - h^m_k(\hat{x}_{k}^i))^{\rm T}
+ H_k^m P_{k}^{i}(H_k^m)^{\rm T}\bigr) \Bigr\} + \frac{n_y}{2}\ln(2\pi) + \frac{1}{2}\ln|S_{k}^{i, m}|\biggr\} .
\end{equation}

For the messages belong to the BP region ($a \in \mathcal{F}_{\text{BP}}$), we have
\begin{equation}\label{eq:mBP}
\begin{split}
m^{\text{BP}}_{f_i^R \rightarrow a^{i,j}_k}(a^{i,j}_k) \propto & \sum_{a_k^{i,0}} \cdots \sum_{a_k^{i,j-1}} \cdots \sum_{a_k^{i,j+1}} \cdots \sum_{a_k^{i,N_{k,E}}} f_i^R \prod_{j_1 = 0, j_1\neq j}^{N_{k,E}} n_{a_k^{i,j_1} \rightarrow f_i^R}(a_k^{i,j_1}).
\end{split}
\end{equation}
Recall that from the frame constraint, target $i$ either produces a measurement $j$ at time $k$ or is missed.
That is, if $a_k^{i,j} = 1$, then $a_k^{i,j_1} = 0, j_1 = 0, \ldots, j-1, j+1, \ldots, N_{k,E}$.
Eq.~(\ref{eq:mBP}) can be rewritten as
\begin{equation}
\begin{split}
m^{\text{BP}}_{f_i^R \rightarrow a^{i,j}_k}(a^{i,j}_k) =\begin{bmatrix}
m^{\text{BP}}_{f_i^R \rightarrow a_k^{i,j}}(0)  \\
m^{\text{BP}}_{f_i^R \rightarrow a_k^{i,j}}(1)
\end{bmatrix}
=
\begin{bmatrix}
\sum\limits_{j_1 = 1(j_1 \neq j)}^{N_{k,E}} n_{a_k^{i,j_1} \rightarrow f_i^R}(1)\prod\limits_{j_2 = 1(j_2 \neq j_1, j)}^{N_{k,E}}n_{a_k^{i,j_2} \rightarrow f_i^R}(0)  \\
\prod\limits_{j_1 = 1(j_1 \neq j)}^{N_{k,E}}n_{a_k^{i,j_1} \rightarrow f_i^R}(0)
\end{bmatrix}.
\end{split}
\end{equation}

In a similar way, the message $m^{\text{BP}}_{f_j^C \rightarrow a^{i,j}_k}(a^{i,j}_k)$ can be rewritten as
\begin{equation}
\begin{split}
m^{\text{BP}}_{f_j^C \rightarrow a^{i,j}_k}(a^{i,j}_k) \propto & \sum_{a_k^{0,j}} \cdots \sum_{a_k^{i-1,j}} \cdots \sum_{a_k^{i+1,j}} \cdots \sum_{a_k^{N_T,j}} f_j^C \prod_{i_1 = 0(i_1\neq i)}^{N_T} n_{a_k^{i_1,j} \rightarrow f_j^C}(a_k^{i_1,j}) \\
= & \begin{bmatrix}
m^{\text{BP}}_{f_j^C \rightarrow a_k^{i,j}}(0)  \\
m^{\text{BP}}_{f_j^C \rightarrow a_k^{i,j}}(1)
\end{bmatrix}
=
\begin{bmatrix}
\sum\limits_{i_1 = 1(i_1 \neq i)}^{N_{T}} n_{a_k^{i_1,j} \rightarrow f_j^C}(1)\prod\limits_{i_2 = 1(i_2 \neq i_1, i)}^{N_T}n_{a_k^{i_2,j} \rightarrow f_j^C}(0)  \\
\prod\limits_{i_1 = 1(i_1 \neq i)}^{N_{T}}n_{a_k^{i_1,j} \rightarrow f_j^C}(0)
\end{bmatrix}.
\end{split}
\end{equation}

According to Eq.~(\ref{eq:bx}), the belief $b_A(a_k^{i,j})$ can be computed as
\begin{equation}\label{eq:baakij}
b_A(a_k^{i,j}) \propto m^{\text{MF}}_{p_{A_k|e_k} \rightarrow a_k^{i,j}}(a_k^{i,j}) \times m^{\text{MF}}_{p_{y^{j}_k|x^{i_1}_k,m_k^{i_1}, a_k^{i,j}} \rightarrow a_k^{i,j}}(a_k^{i,j})  \times m^{\text{BP}}_{f_i^R \rightarrow a^{i,j}_k}(a^{i,j}_k) \times  m^{\text{BP}}_{f_j^C \rightarrow a^{i,j}_k}(a^{i,j}_k).
\end{equation}

Accordingly, the expectation $\hat a_k^{i,j}$ is given by
\begin{equation}\label{eq:eakij}
\begin{split}
\hat a_k^{i,j} = \dfrac{b_A(a_k^{i,j} = 1)}{b_A(a_k^{i,j} = 1) + b_A(a_k^{i,j} = 0)}
               = & \dfrac{1}{1 + \dfrac{ m^{\text{MF}}_{p_{A_k|e_k} \rightarrow a_k^{i,j}}(0) \times m^{\text{MF}}_{p_{y^{j}_k|x^{i}_k,m_k^{i}, a_k^{i,j}} \rightarrow a_k^{i,j}}(0)  \times m^{\text{BP}}_{f_i^R \rightarrow a^{i,j}_k}(0) \times  m^{\text{BP}}_{f_j^C \rightarrow a^{i,j}_k}(0)}{ m^{\text{MF}}_{p_{A_k|e_k} \rightarrow a_k^{i,j}}(1) \times m^{\text{MF}}_{p_{y^{j}_k|x^{i}_k,m_k^{i}, a_k^{i,j}} \rightarrow a_k^{i,j}}(1)  \times m^{\text{BP}}_{f_i^R \rightarrow a^{i,j}_k}(1) \times  m^{\text{BP}}_{f_j^C \rightarrow a^{i,j}_k}(1)}} \\
               = & \dfrac{1}{1 + \exp\left(- \ln \bar m^{\text{MF}}_{p_{A_k|e_k} \rightarrow a_k^{i,j}} - \ln \bar m^{\text{MF}}_{p_{y^{j}_k|x^{i}_k,m_k^{i}, a_k^{i,j}} \rightarrow a_k^{i,j}}- \ln \bar m^{\text{BP}}_{f_i^R \rightarrow a^{i,j}_k} - \ln \bar m^{\text{BP}}_{f_j^C \rightarrow a^{i,j}_k}\right)}
\end{split}
\end{equation}
with
\begin{align}
&\bar m^{\text{MF}}_{p_{A_k|e_k} \rightarrow a^{i,j}_k} = \dfrac{\bar m^{\text{MF}}_{p_{A_k|e_k}\rightarrow a^{i,j}_k}(1)}{\bar m^{\text{MF}}_{p_{A_k|e_k}\rightarrow a^{i,j}_k}(0)} = m^{\text{MF}}_{p_{A_k|e_k} \rightarrow a^{i,j}_k}(1),\label{eq:mMF1} \\
&\bar m^{\text{MF}}_{p_{y^j_k|x^{i}_k,m_k^{i}, a_k^{i,j}\rightarrow a^{i,j}_k}} = \dfrac{ m^{\text{MF}}_{p_{y^{j}_k|x^{i}_k,m_k^{i}, a_k^{i,j}}\rightarrow a^{i,j}_k}(1)}{ m^{\text{MF}}_{p_{y^{j}_k|x^{i}_k,m_k^{i}, a_k^{i,j}}\rightarrow a^{i,j}_k}(0)} = m^{\text{MF}}_{p_{y^j_k|x^{i}_k,m_k^{i}, a_k^{i,j}\rightarrow a^{i,j}_k}}(1),\label{eq:mMF2} \\
&\bar m^{\text{BP}}_{f_i^R \rightarrow a^{i,j}_k} = \dfrac{ m^{\text{BP}}_{f_i^R \rightarrow a^{i,j}_k}(1)}{ m^{\text{BP}}_{f_i^R \rightarrow a^{i,j}_k}(0)}
= \dfrac{\prod\limits_{j_1 = 1(j_1 \neq j)}^{N_{k,E}}n_{a_k^{i,j_1} \rightarrow f_i^R}(0)}{\sum\limits_{j_1 = 0(j_1 \neq j)}^{N_{k,E}} n_{a_k^{i,j_1} \rightarrow f_i^R}(1)\prod\limits_{j_2 = 1(j_2 \neq j)}^{N_{k,E}}n_{a_k^{i,j_2} \rightarrow f_i^R}(0)/n_{a_k^{i,j_1} \rightarrow f_i^R}(0)}, \label{eq:mBP1} \\
&\bar m^{\text{BP}}_{f_j^C \rightarrow a^{i,j}_k} = \dfrac{ m^{\text{BP}}_{f_j^C \rightarrow a^{i,j}_k}(1)}{ m^{\text{BP}}_{f_j^C \rightarrow a^{i,j}_k}(0)}
\label{eq:mBP2}
= \dfrac{\prod\limits_{i_1 = 1(i_1 \neq i)}^{N_{T}}n_{a_k^{i_1,j} \rightarrow f_j^C}(0)}{\sum\limits_{i_1 = 0(i_1 \neq i)}^{N_{T}} n_{a_k^{i_1,j} \rightarrow f_j^C}(1)\prod\limits_{i_2 = 1(i_2 \neq i)}^{N_T}n_{a_k^{i_2,j} \rightarrow f_j^C}(0)/n_{a_k^{i_1,j} \rightarrow f_j^C}(0)}.
\end{align}
Note that the variable to factor messages $n_{a_k^{i,j} \rightarrow f_i^R}(a_k^{i,j})$ and $n_{a_k^{i,j} \rightarrow f_j^C}(a_k^{i,j})$
in Eqs.~(\ref{eq:mBP1}), (\ref{eq:mBP2}) are
\begin{equation}\label{eq:na2fiR}
n_{a_k^{i,j} \rightarrow f_i^R}(a_k^{i,j}) = \begin{cases}
m^{\text{MF}}_{p_{A_k|e_k} \rightarrow a_k^{i,j}}(a_k^{i,0}), & \forall i >0, j = 0 \\
m^{\text{MF}}_{p_{A_k|e_k} \rightarrow a_k^{i,j}}(a_k^{i,j}) m^{\text{MF}}_{p_{y^{j}_k|x^i_k,m_k^i, a_k^{i,j}} \rightarrow a_k^{i,j}}(a_k^{i,j}) m^{\text{BP}}_{f_{j}^C \rightarrow a^{i,j}_k}(a^{i,j}_k), & \forall i >0, j > 0
\end{cases}
\end{equation}
and
\begin{equation}\label{eq:na2fjC}
n_{a_k^{i,j} \rightarrow f_j^C}(a_k^{i,j}) = \begin{cases}
m^{\text{MF}}_{p_{y^{j}_k|x^i_k,m_k^i, a_k^{i,j}} \rightarrow a_k^{i,j}}(a_k^{0,j}), & \forall i = 0, j > 0 \\
m^{\text{MF}}_{p_{A_k|e_k} \rightarrow a_k^{i,j}}(a_k^{i,j}) m^{\text{MF}}_{p_{y^{j}_k|x^i_k,m_k^i, a_k^{i,j}} \rightarrow a_k^{i,j}}(a_k^{i,j}) m^{\text{BP}}_{f_{i}^R \rightarrow a^{i,j}_k}(a^{i,j}_k), & \forall i >0, j > 0
\end{cases}
\end{equation}
Substituting Eqs.~(\ref{eq:na2fiR}), (\ref{eq:na2fjC}) into Eqs.~(\ref{eq:mBP1}), (\ref{eq:mBP2}), yields,
\begin{align}
&\bar m^{\text{BP}}_{f_i^R \rightarrow a^{i,j}_k}(a^{i,j}_k) = \dfrac{1}{m^{\text{MF}}_{p_{A_k|e_k} \rightarrow a_k^{i,0}}(1) + \sum \limits_{j_1 >0(j_1 \neq j)} m^{\text{MF}}_{p_{A_k|e_k} \rightarrow a_k^{i,j_1}}(1) m^{\text{MF}}_{p_{y^{j_1}_k|x^i_k,m_k^i, a_k^{i,j_1}} \rightarrow a_k^{i,j_1}}(1) \bar m^{\text{BP}}_{f_{j_1}^C \rightarrow a^{i,j_1}_k}} \label{eq:mfrakij}, \\
&\bar m^{\text{BP}}_{f_j^C \rightarrow a^{i,j}_k}(a^{i,j}_k) = \dfrac{1}{m^{\text{MF}}_{p_{y^{j}_k|x^i_k,m_k^i, a_k^{i,j}} \rightarrow a_k^{0,j}}(1) + \sum \limits_{i_1 >0(i_1 \neq i)} m^{\text{MF}}_{p_{A_k|e_k} \rightarrow a_k^{i_1,j}}(1) m^{\text{MF}}_{p_{y^{j}_k|x^{i_1}_k,m_k^{i_1}, a_k^{i_1,j}} \rightarrow a_k^{i_1,j}}(1) \bar m^{\text{BP}}_{f_{i_1}^R \rightarrow a^{i_1,j}_k}} \label{eq:mfcakij}.
\end{align}

Note that since loops exist in the BP region when we calculate the belief $b_A(a_k^{i,j})$, the LBP is adopted.
By showing that the message update equations are contractions, the LBP for the data association problem was proved to converge~\cite{williams2014approximate}. By observing Eqs.~(20), (21), (61), (62) in~\cite{williams2014approximate},
the convergence of the LBP in this part is assured by leveraging the techniques in~\cite{williams2014approximate}.

\subsection{Summary}
The proposed MP-MMTT algorithm which solves MMTT based on MP, performs target kinematic state estimation, target visibility state estimation, target motion mode-model decision and data association decision jointly in a closed-loop iterative manner, and is summarized as \textbf{Algorithm}~\ref{alg}.
Like MHT, MP-MMTT works in an online fashion using a sliding window.
\begin{algorithm}
\caption{MP-MMTT algorithm}
\begin{algorithmic}[1]
        \REQUIRE Sequence of measurements $Y_{k - l + 1 : k}, k \geq l$ with $l > 0$ being the interval length;
        \ENSURE Beliefs $b_X(X_{k-l+1:k})$, $b_E(E_{k-l+1:k})$, $b_M(M_{k-l+1:k})$, $b_A(A_{k-l+1:k})$;
        \STATE \underline{\textbf{Initialization:}} initialize beliefs $b_i^{(0)}(x_i)$ for all $i \in \mathcal{I}_{\text{MF}} \backslash \mathcal{I}_{\text{BP}}$, i.e., $b_X^{(0)}(X_{1:K})$, $b_E^{(0)}(E_{1:K})$, $b_M^{(0)}(M_{1:K})$ and $N_T$; send the corresponding messages $n_{i \rightarrow a}(x_i) = b_i^{0}(x_i)$ to all factor nodes $a \in \mathcal{S}_{\text{MF}}(i)$.
        \FOR{each iteration}
        \STATE  \underline{\textbf{Data association:}} Calculate the belief $b_A(a)$ and its expectation $\hat a_k^{i,j}$ iteratively via
        Eqs.~(\ref{eq:baakij}), (\ref{eq:eakij}) by using all incoming messages given by Eqs.~(\ref{eq:mfrakij}), (\ref{eq:mfcakij}).
        \STATE \underline{\textbf{Visibility state estimation:}} Calculate the belief $b_E(e)$ via Eq.~(\ref{eq:bee1k}) by using a forward~backward algorithm.
        \STATE \underline{\textbf{Mode-model association:}} Calculate the belief $b_M(m)$ via Eq.~(\ref{eq:bMm}) by using a forward~backward algorithm.
        \STATE  \underline{\textbf{Kinematic state estimation:}} Calculate the mode-dependent beliefs $b_X(x^{m})$ via a fixed interval smoother, and then fuse those mode-dependent beliefs to obtain the belief $b_X(x)$ via Eq.~(\ref{eq:bx1k}).
        \STATE \underline{\textbf{Iteration stop rule:}} the iteration terminates if the beliefs between two consecutive iterations are close enough~(less than the iteration threshold $\delta_T$) or the maximum number of iterations $r_{\text{max}}$ is reached.
        \ENDFOR
        \STATE Perform track management by using the visibility probability $b_E(e)$.
        \STATE Exit the iteration and go to the next sliding window.
\end{algorithmic}
\label{alg}
\end{algorithm}

\subsection{Other Aspects of MP-MMTT}
\subsubsection{Initialization}
Initial beliefs $b_X^{(0)}(X_{1:l})$, $b_E^{(0)}(E_{1:l})$, $b_M^{(0)}(M_{1:l})$ and the maximum number of targets $N_T$ are required for MP-MMTT.
The initialization procedure for the first sliding window $k \in [1, l]$ is given as follows.
\begin{itemize}
\item At time $k = 1$, the tentative tracks are established for each measurement, i.e., each measurement is an ``initiator''.
\item At time $k = 2$, a gate associated with each initiator is set up based on the assumed maximum target velocity and the measurement noise intensity. If a measurement falls in the gate, then the associated tentative track becomes a preliminary track. Otherwise, the tentative track is dropped. For each preliminary track that has two measurements, a filter can be used to initialize the kinematic state estimate $\{\hat x_{2}^i, P_{2}^i\}$. Set up the suitable values of initial visibility probability $\pi_{i, e}$ and initial model probability $\pi_{i, m}$.
\item Starting from $k = 3$, for each preliminary track $i$, select candidate measurements using gating technique, establish the pseudo-measurement via the LBP algorithm, and update the kinematic state $\{\hat x_{k}^i, P_{k}^i\}$ by a filter. Meanwhile, the visibility probability $p(e_{k}^i)$ and model probability $p(m_{k}^i)$ are recursively updated by using forward algorithm. The measurements that do not fall into the validation gates of any tracks are used to initialize new tracks.
\item For the batch window $[1, l]$, manage tracks based on $p(E_{1:l})$. Specifically, if the average visibility probability of target $i$ in three successive scans is less than the threshold $\delta_d$, the track $i$ is deleted; otherwise, track $i$ becomes a confirmed track. $N_{T}$ is the total number of confirmed potential tracks over the batch window $[1, l]$.
\end{itemize}

\subsubsection{Computational Complexity}
The MP-MMTT algorithm is an iterative processor with MP among four subgraphs.
Its computational complexity is
\begin{equation}
c_{total} = N_r \times (c_x + c_e + c_m + c_a),
\end{equation}
where $N_r$ is the number of iterations, $c_x$, $c_e$, $c_m$ and $c_a$ are the computational cost of target kinematic state estimation, target visibility state estimation, target motion mode-model association and data association, respectively. For target kinematic state estimation, the main cost relies on the model-dependent state estimation, which is carried out by a fixed-interval smoother. If Kalman smoother is used, $c_x = \mathcal{O}(lN_Tn_x^3)$. Both the target visibility state estimation and target motion model association are carried out by forward and backward algorithm, thus $c_e = \mathcal{O}(4lN_T)$ and $c_m = \mathcal{O}(lN_TN_M^2)$. LBP is used to approximate the data association. The main computational cost of LBP is the message update equation, which is $\mathcal{O}(|x_i|^2)$ for each variable node $i$ at each iteration.
In the data association subgraph, there are at most $(N_T + 1)(N_{k, M} + 1)$ variable nodes, each of which takes values of 0 and 1. Hence, the computational cost $c_{a} = \mathcal{O}(N_{a}\sum_{k=1}^{l}4N_TN_{k, E})$ with $N_{a}$ being the number of LBP iterations.

%\subsubsection{Convergence Analysis}
%\textcolor{blue}{As stated in~\cite{riegler2012merging}, if the BP region has no cycle and the following inequality is fulfilled
%\begin{equation}
%  |\mathcal{N}(a) \cap \mathcal{I}_{BP}| \leq 1, \quad \forall a \in \mathcal{F}_{MF}
%\end{equation}
%then the combined BP-MF approach is guaranteed to converge, by the fact that running the message update equation in the BP part cannot increase the region-based free energy approximation $F_{BP, MF}$ in Eq.(8). If the factor graph representing the BP region contains loops, i.e., not cycle free, then the BP-MF approach is not guaranteed to converge since there is no guarantee that LBP will converge for general graphs with loops.
%In our work, the inequality is fulfilled by the fact that $|\mathcal{N}(a) \cap \mathcal{I}_{BP}| = |A| = 1$. Although the BP region contains the loops among data association events, the LBP for iterative message update equation converges. \textbf{In this case, the convergence of BP-MF approach is still an open problem, and this will be our further work.}
%}

\subsubsection{Properties of MP-MMTT}
The proposed MP-MMTT algorithm has several properties:
\begin{itemize}
\item It is obtained by a unified MP approach that performs the BP and MF approximation for MMTT.
\item It has a closed-loop iterative manner among kinematic state estimation, target detection, data association decision, and motion mode-model association decision. In the view of feedback control, such an iterative structure of MP-MMTT is effective in dealing with the coupling between estimation error and decision error.
\item It is computationally effective. Leveraging MF approximation, the joint inference of high-dimensional latent variables is decomposed into several individual inferences of low-dimensional latent variables. Meanwhile, the problem of the combinatorial explosion in data association is eliminated by using LBP.
\end{itemize}

\section{Simulation and Analysis}\label{sec:simulation}
We consider a simulation scenario with an unknown and time-varying number of maneuvering targets in the presence of clutter and missed detections.
The proposed MP-MMTT algorithm is compared with IMMJPDA~\cite{yaakov1992}, IMMHMHT~\cite{cox1996, MHT2004} and MMGLMB~\cite{punchihewa2016generalized}.
All the four algorithms are implemented in MATLAB R2016a on a PC with an Intel Core i5 CPU and 8GB RAM.

\emph {1) Scenario parameters}: The surveillance region is assumed to be $[13, 19]~\text{km}$ in range and $[0.7, 1]~\text{rad}$ in azimuth.
Sampling period $T = 1\text{s}$.
Two motion models, constant velocity~(CV) and constant turning~(CT), are selected to model target motion.
The corresponding parameters are
\begin{equation}
F_k^{\text{CV}} = \textrm{I}_2 \otimes \begin{bmatrix}
1 & T  \\
0 & 1
\end{bmatrix},
Q_k^{\text{CV}} = \textrm{I}_2 \otimes \begin{bmatrix}
0.01 & 0 \\
0 & 0.005
\end{bmatrix},
F_k^{\text{CT}} = \begin{bmatrix}
1 & \dfrac{\sin (\theta)}{\omega} & 0 & \dfrac{\cos (\theta)- 1}{\omega} \\
0 & \cos (\theta) & 0 & -\sin (\theta) \\
0 & \dfrac{1 - \cos (\theta)}{\omega} & 1 & \dfrac{\sin(\theta)}{\omega} \\
0 & \sin(\theta)& 0 & \cos(\theta)
\end{bmatrix}
, Q_k^{\text{CT}} = 10 Q_k^{\text{CV}}.
\end{equation}
where $\omega = 0.087~\text{rad}$ and $\theta = \omega T$.
The measurement noise covariance $R = \text{diag}(400~\text{m}^2, 1e-6~\text{rad}^2)$.
The initial kinematic states and motion parameters of four targets are shown in Table~\ref{targets}.
\begin{table}[!htbp] \renewcommand\arraystretch{1.2}
   \centering
    \caption{\label{targets} The initial state and motion parameters of targets}
    \begin{tabular}{c|c|c|c|c|c}
        \hline
        \multirow{2}{*}{\textbf{target index $i$}} & \multirow{2}{*}{\textbf{{Initial kinematic state}}} & \multirow{2}{*}{\textbf{{Duration}}} & \multicolumn{3} {|c} {\textbf{{Motion model and its duration}}}  \\ \cline{4-6}
         & & &  CV & CT & CV \\ \hline
        $i = 1$ & [11400; 0; 10200; 120]  & [1,30]  & [1,10]  & [11,20] & [21,30]  \\ \hline
        $i = 2$ & [11300; 0; 10200; 120]  & [1,30]  & [1,10]  & [11,20] & [21,30] \\ \hline
        $i = 3$ & [11750; -120; 11840; 0] & [11,40] & [11,20] & [21,30] & [31,40]  \\ \hline
        $i = 4$ & [11750; -120; 11940; 0] & [11,40] & [11,20] & [21,30] & [31,40] \\ \hline
    \end{tabular}
\end{table}

The performance of target tracking algorithms is related to detection probability $P_d$, clutter density $\lambda$, the number of targets $N_T$ and the distance between parallel-moving targets.
One hundred Monte Carlo runs are carried out to compare the performances of MP-MMTT, IMMJPDA, IMMHMHT, and MMGLMB by varying these factors.

\emph {2) Algorithm parameters}: For MP-MMTT, gate probability $P_g = 0.997$, threshold for terminating the iteration $\delta_T = 10^{-3}$, maximum number of iterations $r_{max} = 10$, window length $l=10$ and sliding step $s = 1$.
The thresholds of survival target and dead target are $\delta_c = 0.85$ and $\delta_d = 0.3$, respectively.
The initial visibility probability for each target $\pi_{i,e=1} = 0.5$.
The detection probability related to the target visibility state $P_d^i(1) = 0.9$ and $P_d^i(0) = 0.1$.
A track is terminated if $p(s_k^i = 1)$ is less than $\delta_s$ for three successive scans.
Let the initial model probability for each target $\pi_{i,m = \text{CV}} = 0.9$, $\pi_{i,m = \text{CT}} = 0.1$, and
the transition probability matrices
$
T_{i,e} = \begin{bmatrix}
\begin{matrix}
 0.85 & 0.15 \\
 0.15 & 0.85
\end{matrix}
\end{bmatrix}
,
T_{i,m} =
\begin{bmatrix}
\begin{matrix}
0.9 & 0.1 \\
0.1 & 0.9
\end{matrix}
\end{bmatrix}$.
For IMMJPDA, M/N logic rule with parameters $2/2\&1/3$ is used for track confirmation.
Specifically, a new track is confirmed if at least three detections are received over five successive scans, and the first two detections are used to initialize the track head.
Based on the hypothesis-oriented MHT algorithm, IMMHMHT is implemented by adopting Murty's approximation method that
always keeps the first $n$~($n=3$ in this paper) best hypothesis, and the window length is 3.
For both IMMJPDA and IMMHMHT, a track will be deleted if no measurement falls into the gate of the track over three successive scans.
To improve the computational efficiency and reduce the number of false tracks,
GLMB filters often assume the locations that new targets possibly appear are known to be in a small set~\cite{ristic2012adaptive}.
For MMGLMB, the four targets are assumed to appear at four pre-defined Gaussian birth locations with probability $0.025$ and means being $[11400, 0, 10200, 0]^T$, $[11300, 0, 10200, 0]^T$,  $[11750, 0, 11840, 0]^T$, $[11750, 0, 11940, 0]^T$, respectively.
The GLMB filter~\cite{Vo2013Labeled} is capped to $100$ components for the four targets.
To compare MMGLMB with IMMJPDA, IMMHMHT, and MP-MMTT, a track is terminated if the label of the track is missed over three successive scans.
Note that the initial model probability and transition probability matrices are set to be the same for the four algorithms.

\emph {3) Performance evaluation}:
The following performance metrics are used to evaluate the four algorithms.
\begin{itemize}
    \setlength{\itemsep}{0pt}
    \setlength{\parsep}{0pt}
    \setlength{\parskip}{0pt}
    \item Number of Valid Tracks (NVT $\uparrow$): A track is valid if it is assigned to only one target and,
 the assigned target is not associated with any other tracks.
    \item Track Probability of Detection (TPD $\uparrow$): Ratio of the length of a valid track to the lifetime of its associated target.
    \item Number of False Tracks (NFT $\downarrow$): A track is false if it is not associated with any target.
  %  \item {Tentative Track Latency (TTL $\downarrow$)}
    \item {Euclidean Error (EE $\downarrow$)}: Euclidean error is defined as the absolute value of the difference between the true value and the estimated value.
    \item Number of Track Breakages~(NTB$\downarrow$): For a target, NTB is defined as one less the number of tracks associated with the target.
    \item {Optimal Subpattern Assignment (OSPA $\downarrow$)}~\cite{Schuhmacher2008A}: A weighted sum performance index considering both detection performance~(measured by cardinality distance) and estimation performance~(measured by spatial distance).
    \item {Mode-model Association Error Rate (MAER $\downarrow$)}: False posterior probability of the model associated with the true mode of the target.
    \item {Data Association Error Rate (DAER $\downarrow$)}: False posterior probability of a valid track associated with the measurement originated from the target.
    \item Total Execution Time (TET $\downarrow$).
\end{itemize}
For the detailed definition of NVT, TPD, and TET, refer to \cite{Gorji2011Performance}.
We denote AOSPA~($\downarrow$) as the averaged OSPA over time.
To statistically evaluate the performance of the four algorithms, TPD, EE, NTB, OSPA, MAER, DAER are averaged overall targets, and all of the metrics are averaged over all Monte Carlo runs.
$\uparrow$~($\downarrow$) indicates the higher~(lower) value the metric,  the better~(worse) the performance is.

%The notations $\uparrow$ indicates that higher scores are better and $\downarrow$ means the opposite.

\emph {4) Simulation results}:
A challenging scenario of four target trajectories is designed.
Target 1 and Target 2 move in parallel along Y-direction with 100~m away in X-direction, and Target 3 and Target 4 move in parallel along X-direction with 100~m away in Y-direction.
Target 1 and Target 3 cross at time $k = 15$, and Target 2 and Target 4 cross at time $k = 16$.
Target 1 and Target 2 maneuver during the time interval [11, 20], and Target 3 and Target 4 maneuver during the time interval [21, 30], respectively. In this scenario, $P_d = 0.95$ and $\lambda = 10^{-4}$.

\begin{figure}[!htbp]
    \centering
    \subfloat{\label{fig12-a}\includegraphics[scale = 0.5]{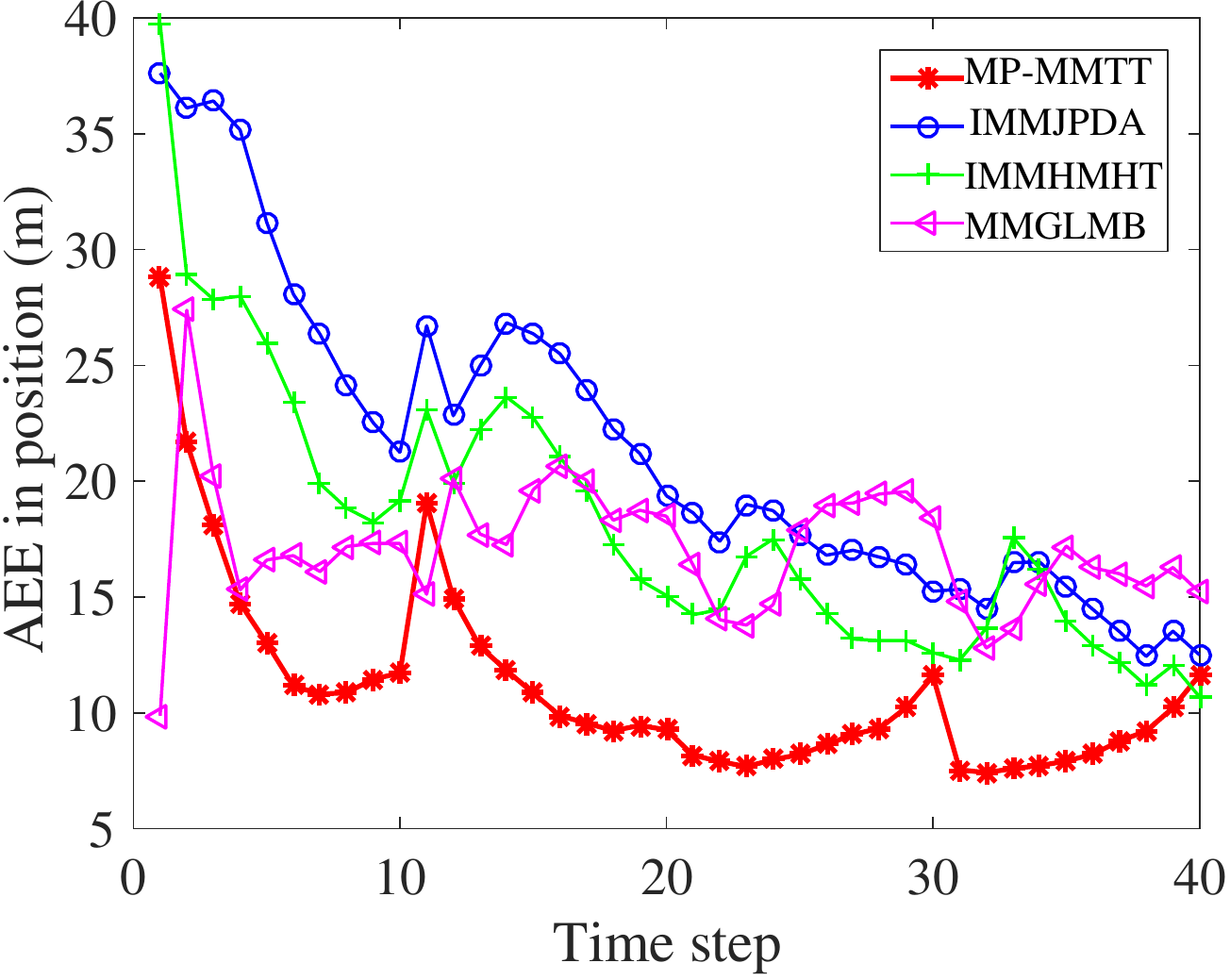}} \hspace{40pt}
    \subfloat{\label{fig12-b}\includegraphics[scale = 0.5]{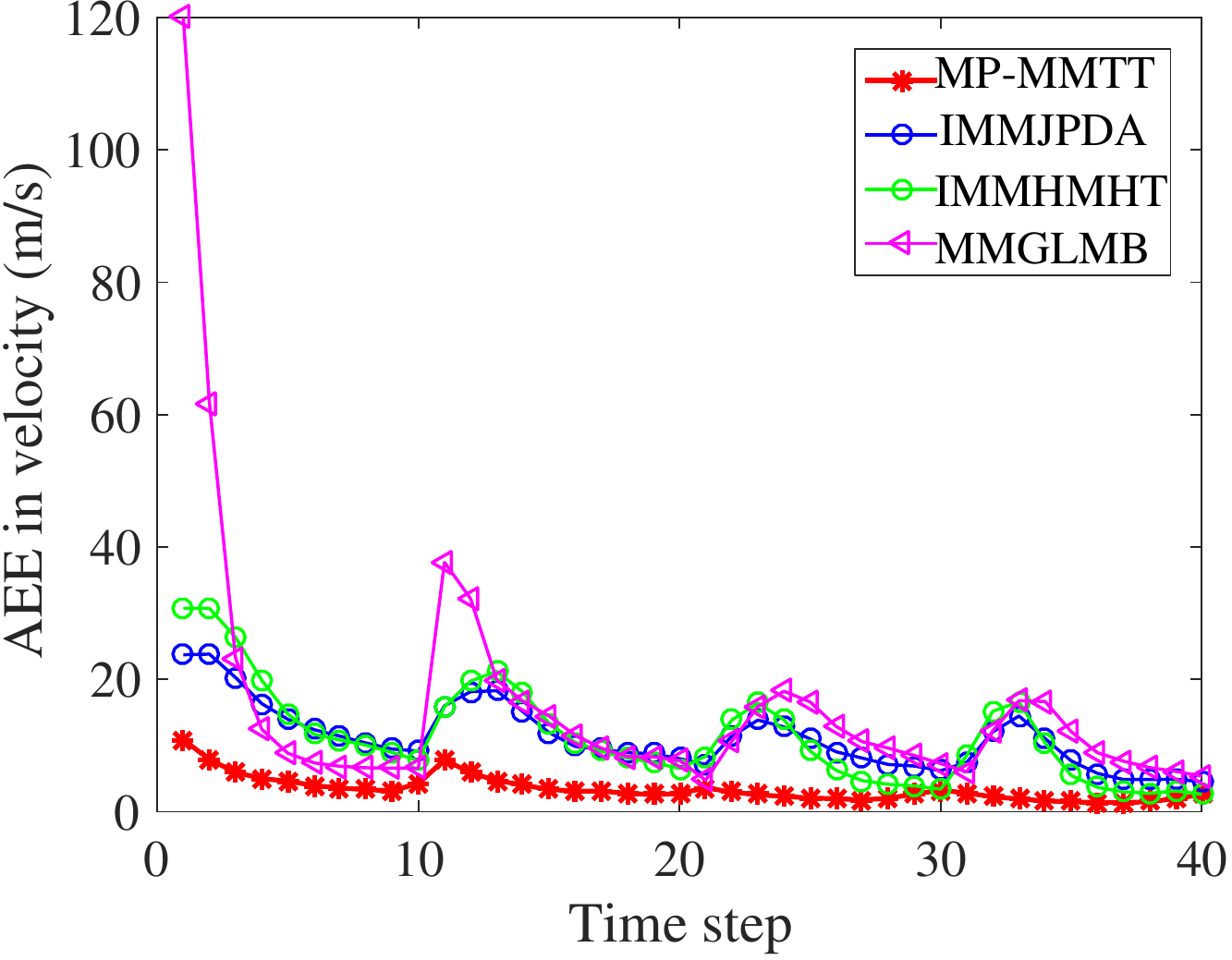}}
    \caption {Target kinematic state estimation} \label{AEE}
\end{figure}

\begin{figure}[!htbp]
\begin{minipage}[t]{0.5\linewidth}
    \centering
    \includegraphics[scale = 0.5]{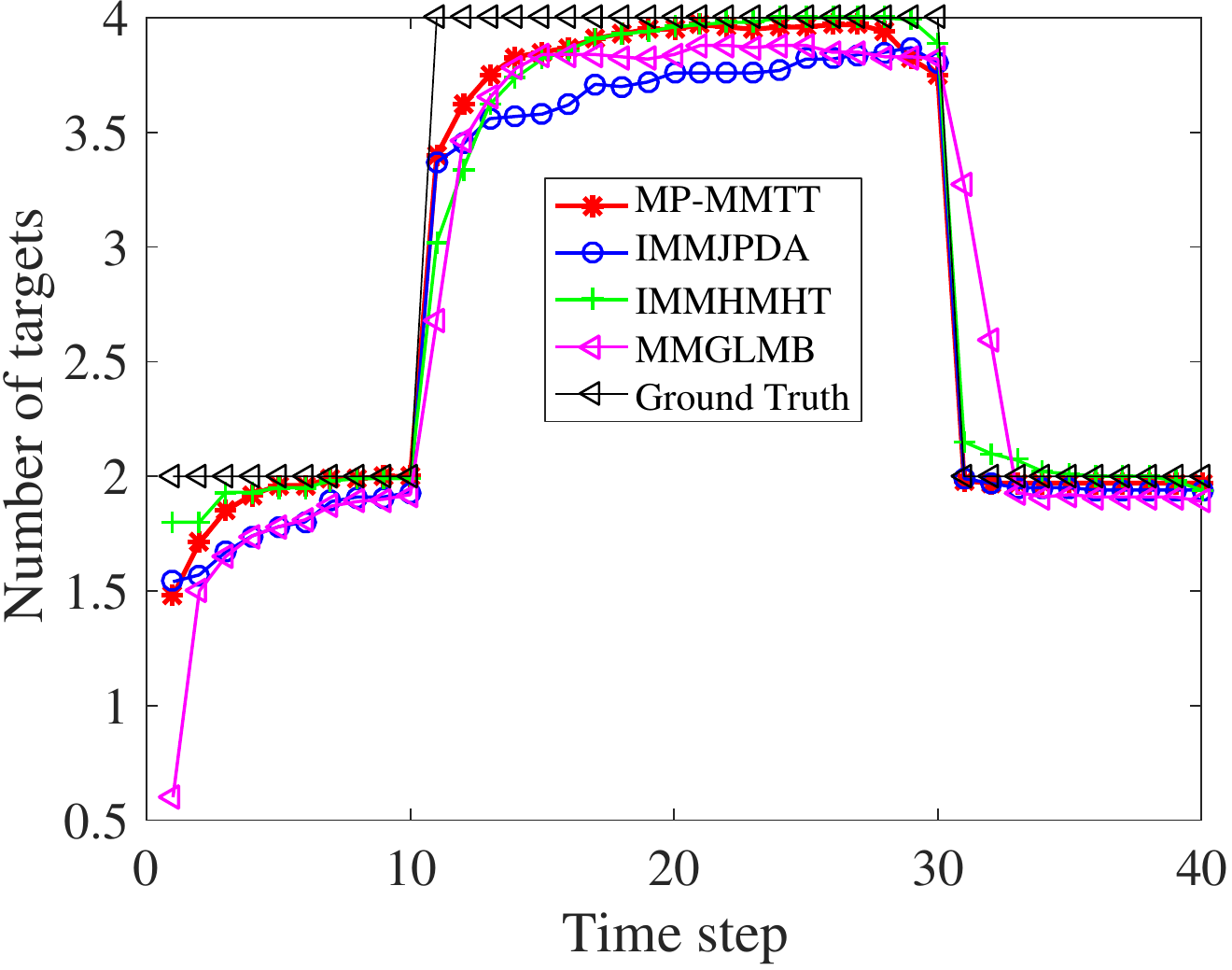}
    \caption{Target visibility state estimation}\label{TargetNum}
\end{minipage}
\begin{minipage}[t]{0.5\linewidth}
    \centering
    \includegraphics[scale = 0.5]{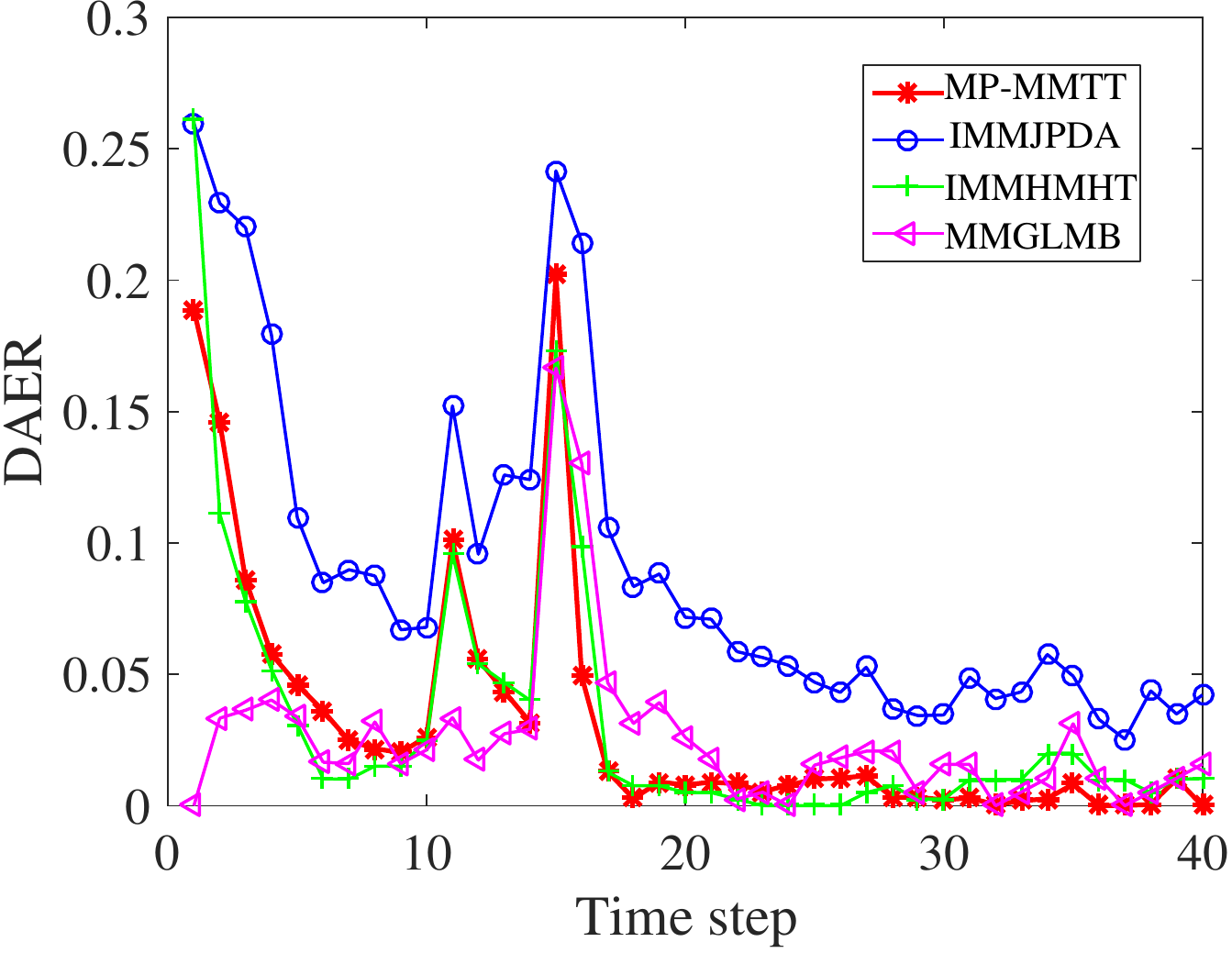}
\caption{Data association error rate}\label{DAER}
\end{minipage}
\end{figure}

\begin{figure}[!htbp]
\begin{minipage}[t]{0.5\linewidth}
    \centering
    \includegraphics[scale = 0.5]{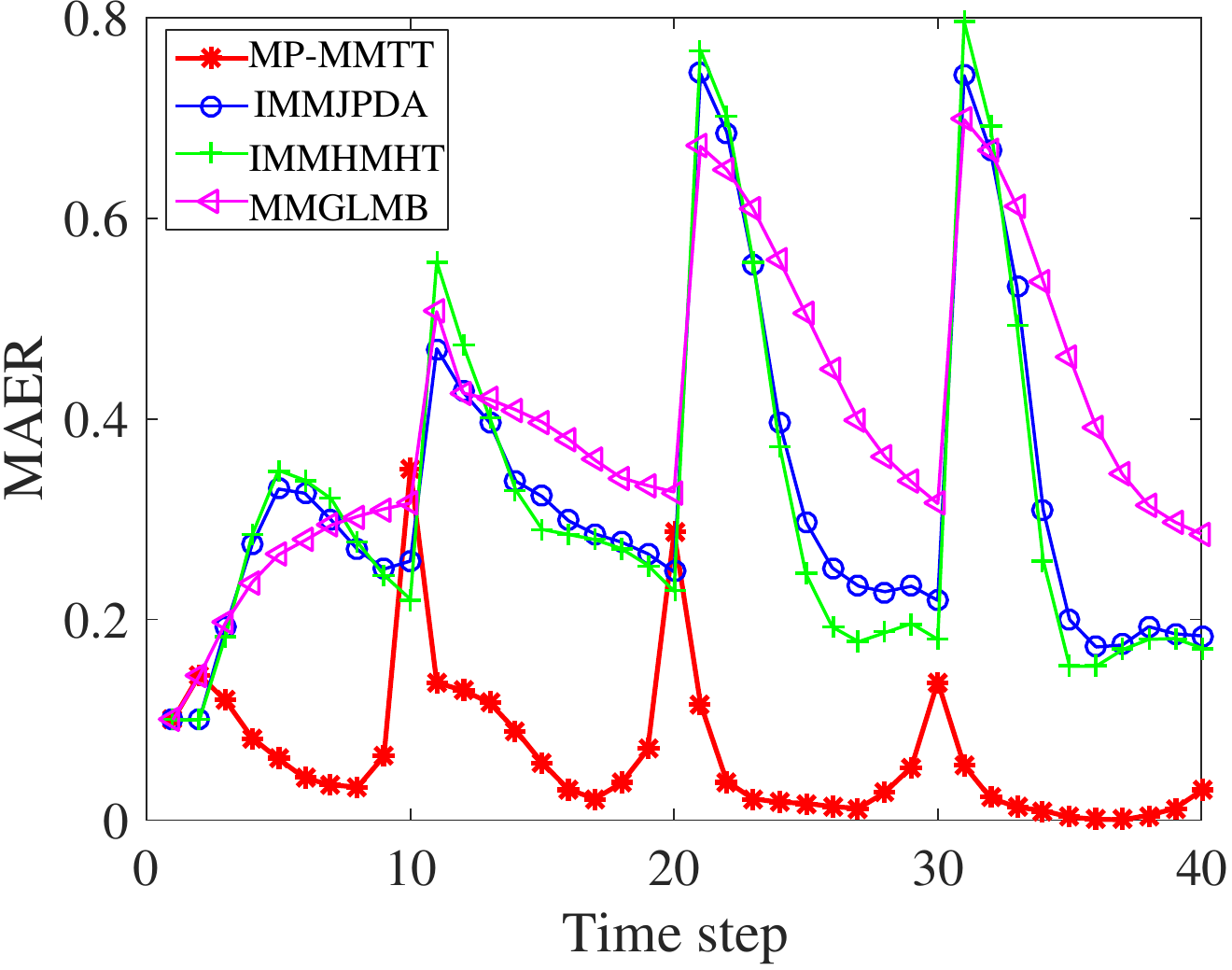}
    \caption{Mode-model association error rate}\label{MAER}
\end{minipage}
\begin{minipage}[t]{0.45\linewidth}
    \centering
    \includegraphics[scale = 0.5]{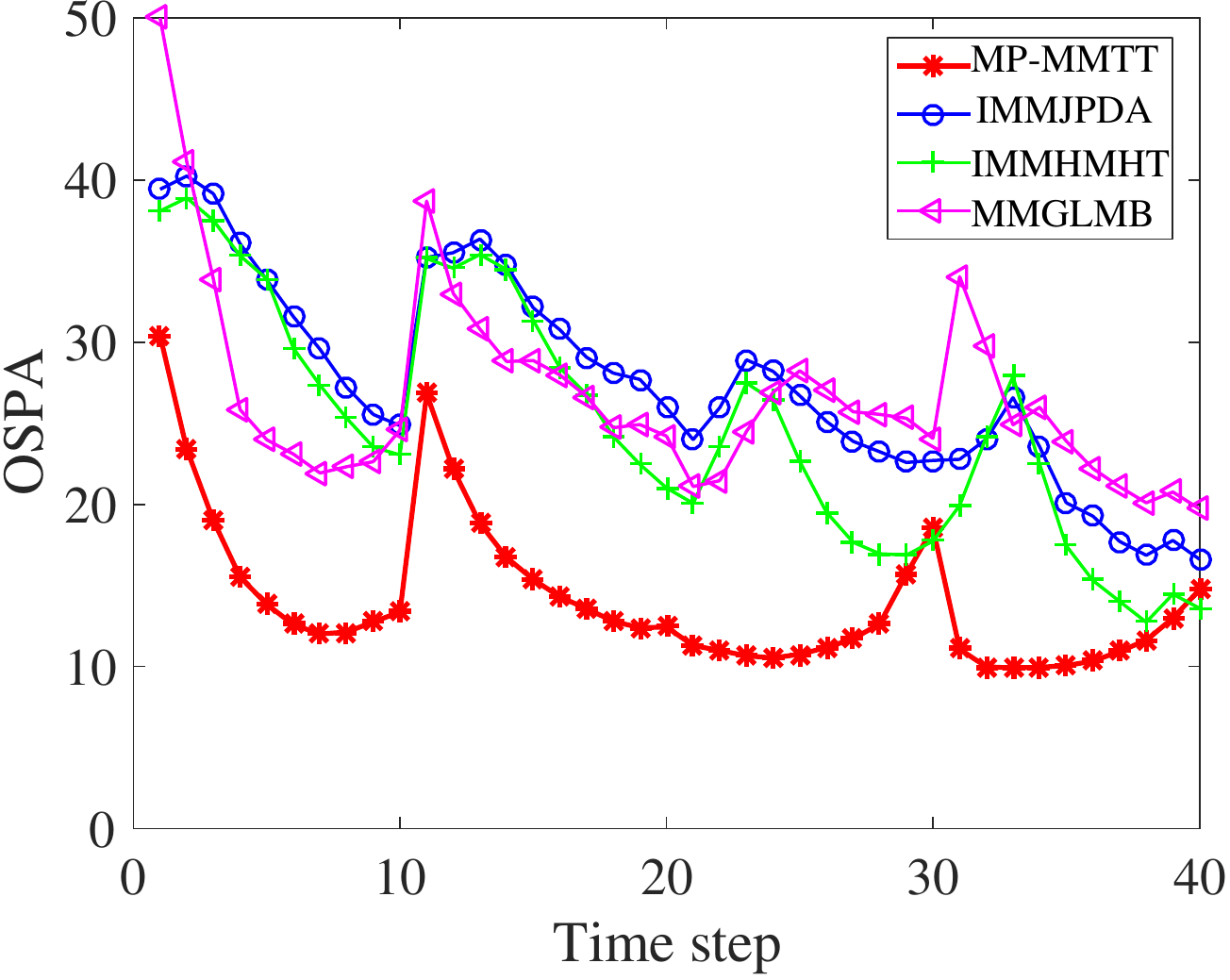}
\caption{OSPA}\label{OSPA}
\end{minipage}
\end{figure}

The performance comparison on target kinematic state estimation, target visibility state estimation, data association, and motion mode-model association of all algorithms are shown in Fig.~\ref{AEE}-Fig.~\ref{MAER}, respectively. MP-MMTT performs best on target state estimation and motion mode-model association.
MP-MMTT and IMMHMHT have comparable performance on estimating the number of targets, which are better than IMMJPDA and MMGLMB. In terms of the data association, IMMHMHT is slightly better than MP-MMTT and MMGLMB, and IMMJPDA is worst. The OSPA in Fig.~\ref{OSPA} shows that, on the whole, MP-MMTT is superior to the other algorithms. The reason is that the iterative and batch processing manner of MP-MMTT is benefit to improve the performance on both estimation and decision.
Note that peaks appear in MAER and OSPA curves at time 10, 20, and 30 due to the birth of new targets and model switch of targets motion.

\begin{figure}[!htbp]
\begin{minipage}[t]{0.5\linewidth}
    \centering
    \includegraphics[scale = 0.5]{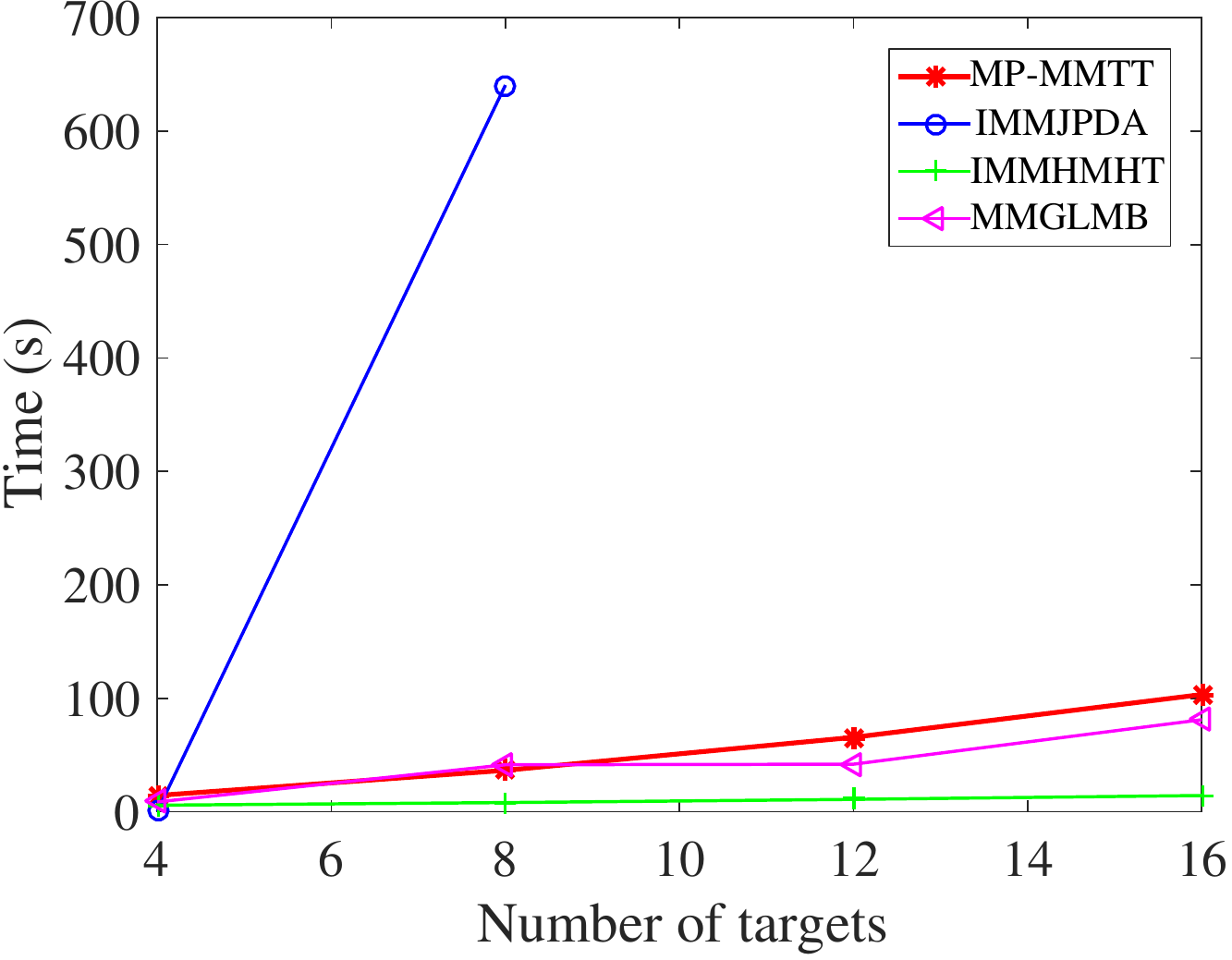}
    \caption{Running time w.r.t. number of targets}\label{Time-num}
\end{minipage}
\begin{minipage}[t]{0.5\linewidth}
    \centering
    \includegraphics[scale = 0.5]{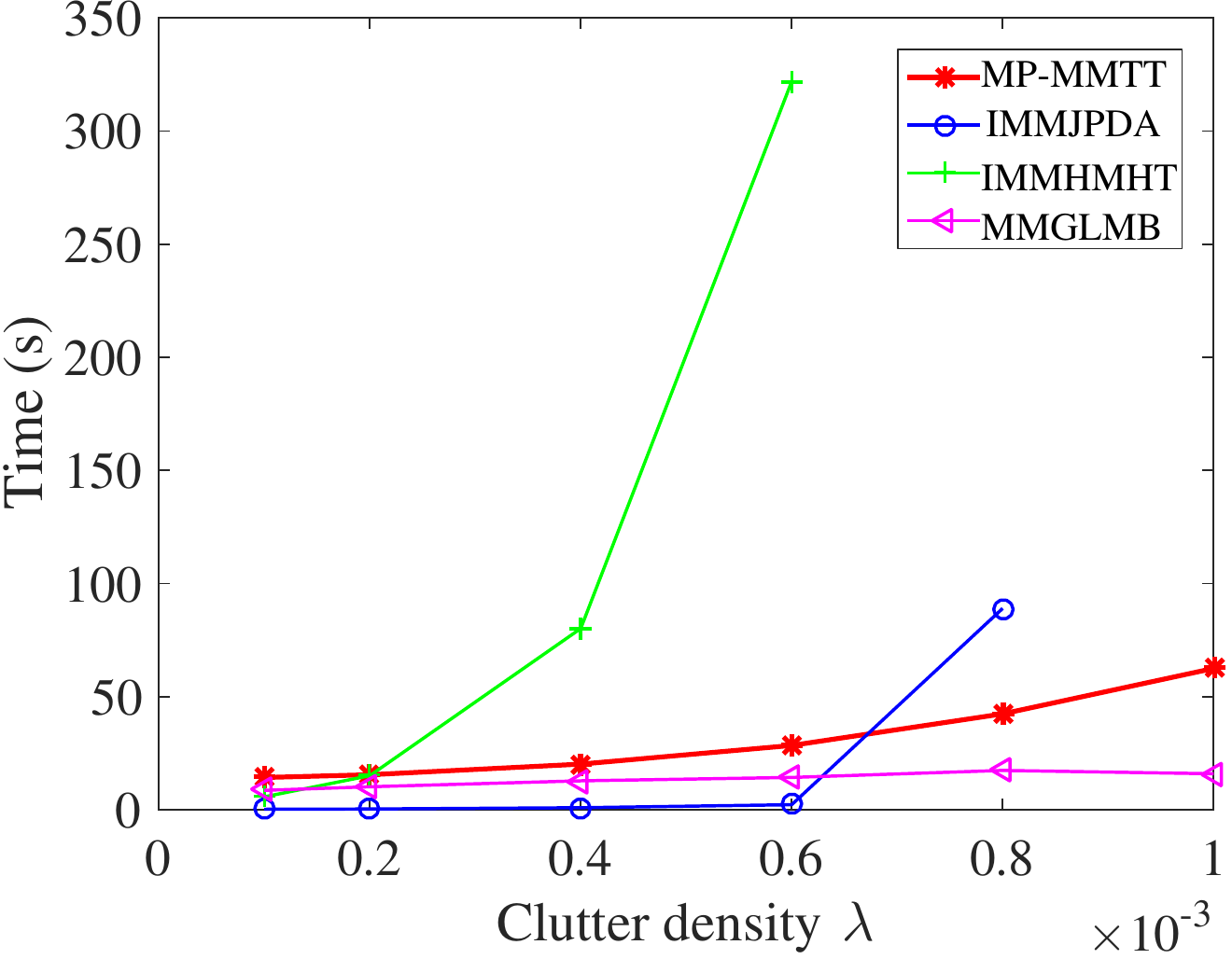}
    \caption{Running time w.r.t. clutter density}\label{Time-pf}
\end{minipage}
\end{figure}

\begin{figure}[!htbp]
\begin{minipage}[t]{0.5\linewidth}
    \centering
    \includegraphics[scale = 0.5]{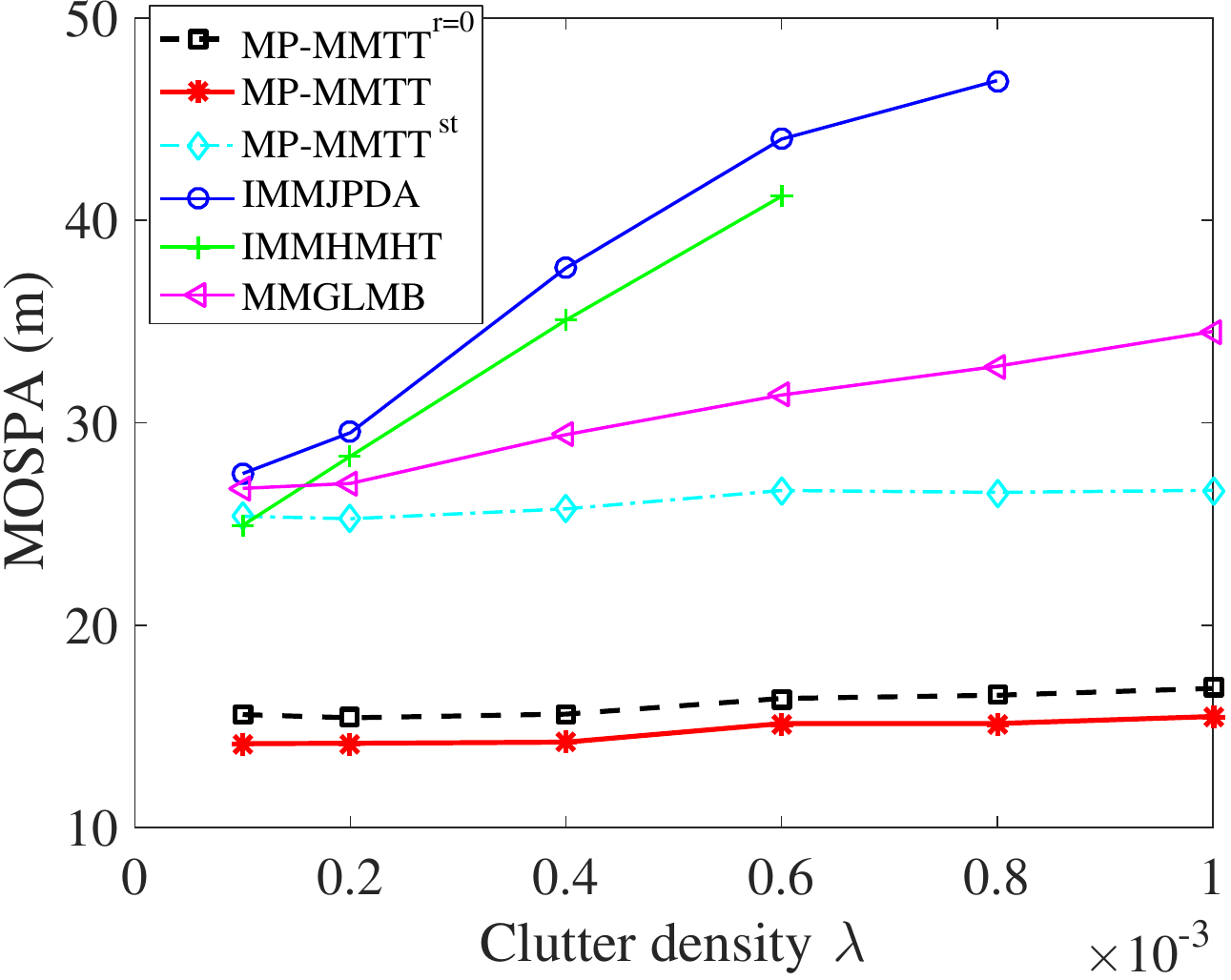}
    \caption{MOSPA w.r.t. clutter density}\label{OSPA-pf}
\end{minipage}
\begin{minipage}[t]{0.5\linewidth}
    \centering
    \includegraphics[scale = 0.5]{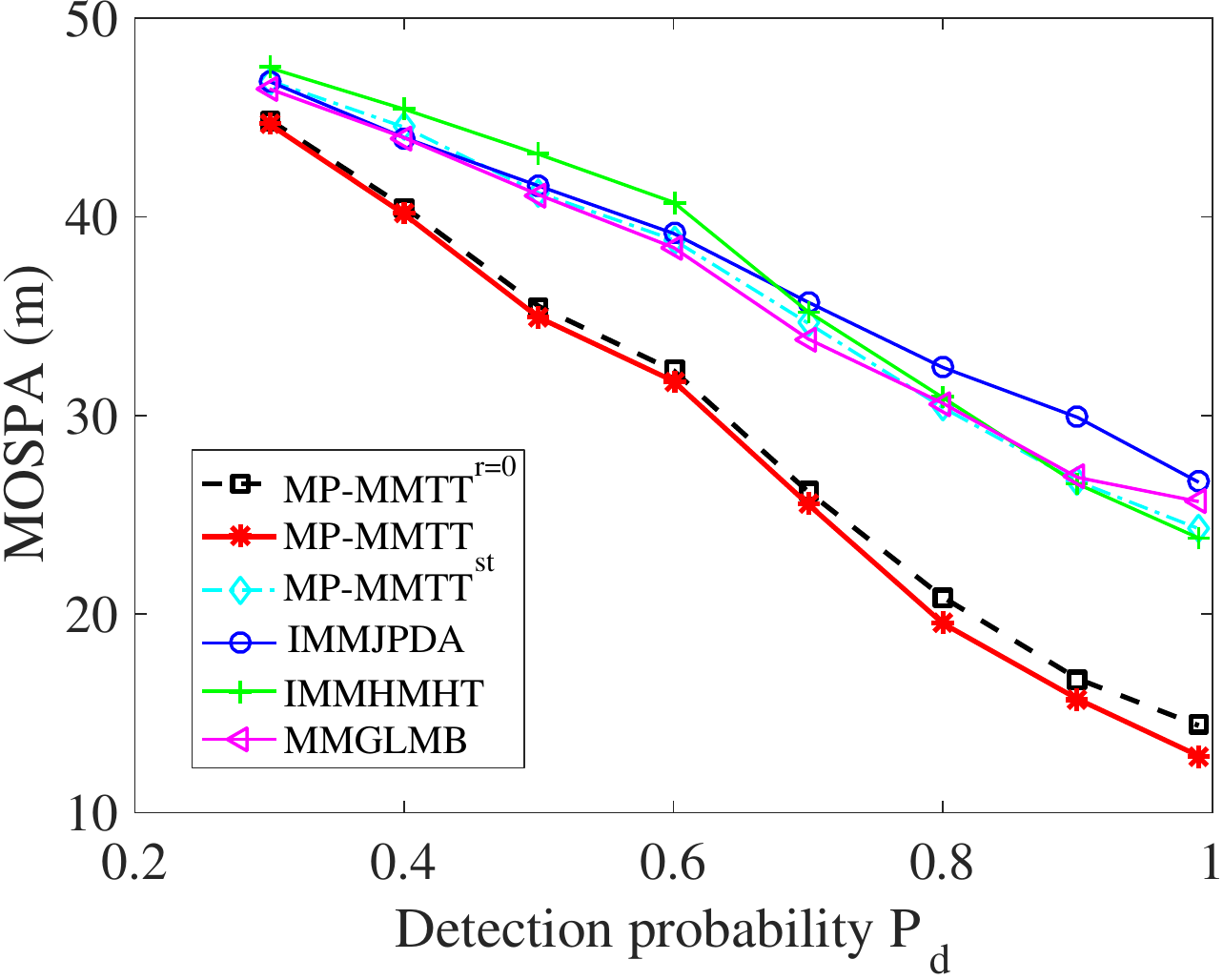}
    \caption{MOSPA w.r.t. detection probability}\label{OSPA-pd}
\end{minipage}
\end{figure}

\begin{figure}[!htbp]
\begin{minipage}[t]{0.5\linewidth}
    \centering
    \includegraphics[scale = 0.5]{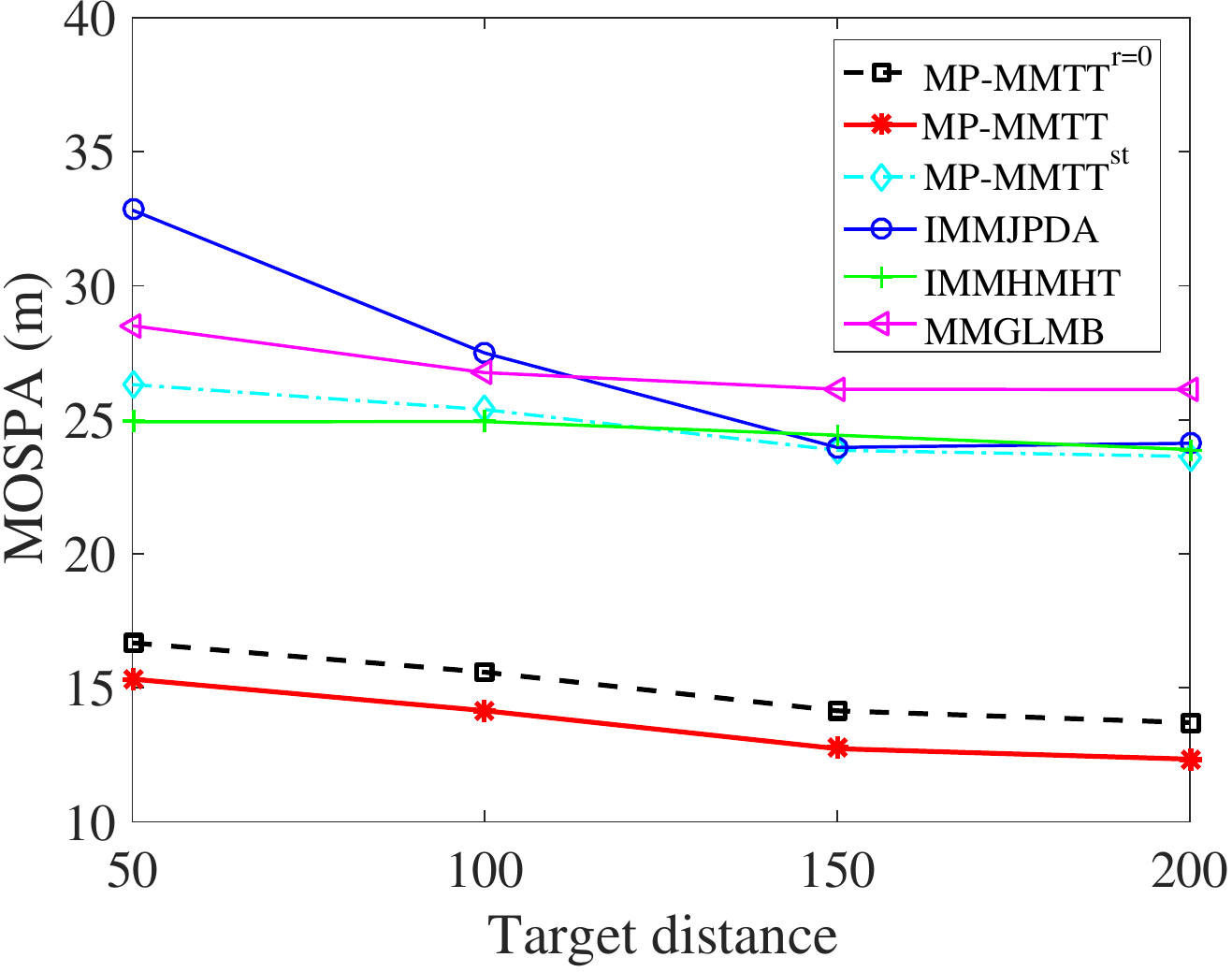}
    \caption{MOSPA w.r.t. target distance}\label{OSPA-space}
\end{minipage}
\begin{minipage}[t]{0.5\linewidth}
    \centering
    \includegraphics[scale = 0.5]{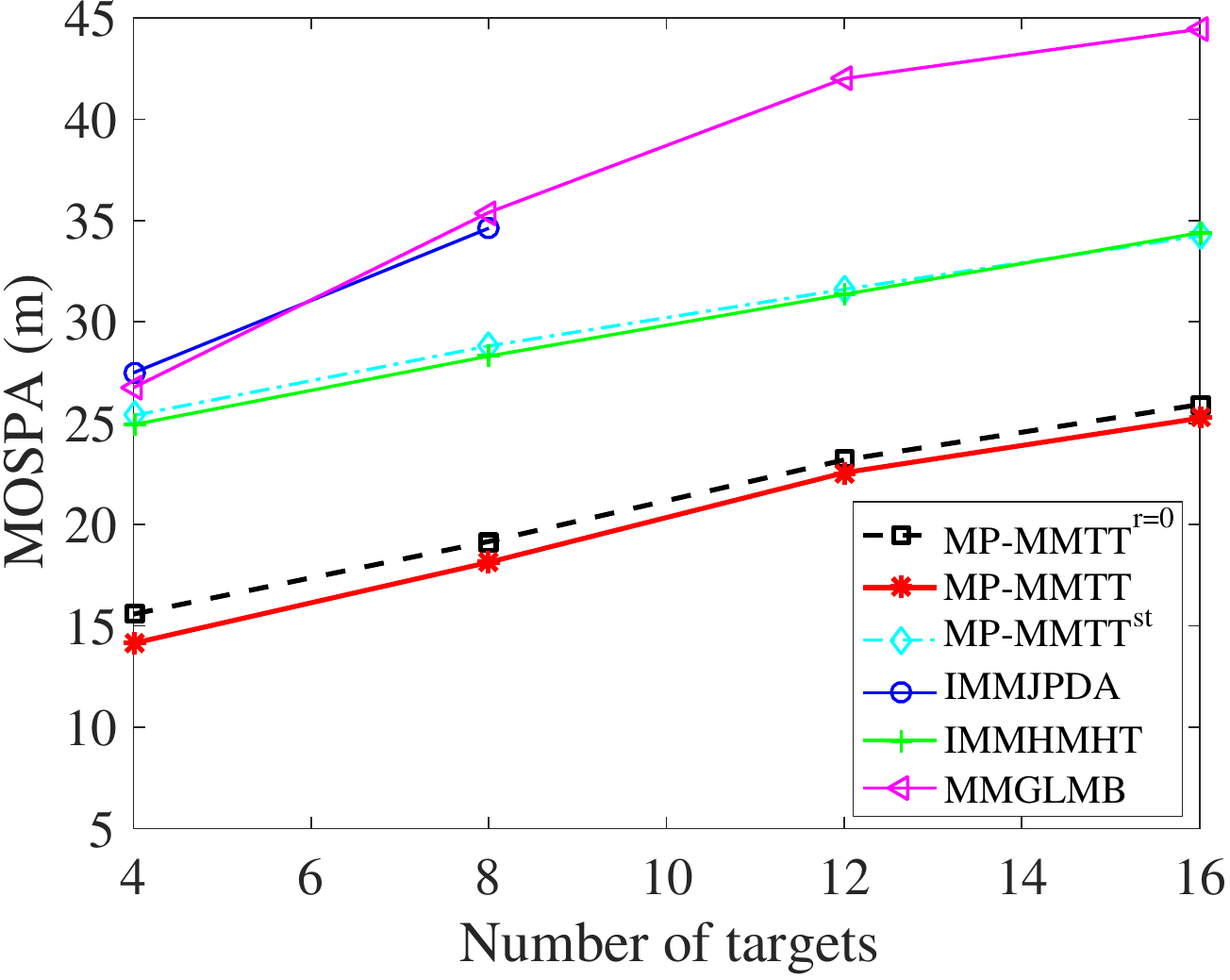}
    \caption{MOSPA w.r.t. number of targets}\label{OSPA-number}
\end{minipage}
\end{figure}

\begin{table}[!htbp]
    \renewcommand \arraystretch{1.2}
    \centering
    \caption{\label{ComPd} Performance Comparison}
    \begin{threeparttable}
    \begin{tabular}{c|c|c|c|c}
        \hline \hline
        \textbf{Metrics}  & \tabincell{c}{IMMJPDA}&{IMMHMHT}& {MMGLMB}& {MP-MMTT} \\ \hline
            NVT  &3.96 &\bf{4} &3.92 &\bf{4}   \\ \hline
            TPD  &0.94 &\bf{0.96} &0.86 &0.95   \\ \hline
            NFT  &0.47 &0.42 &0.26 &\bf{0.02}  \\ \hline
            ANBT  &\bf{0.01} &\bf{0.01} &0.14 &\bf{0.01}  \\ \hline
            MAER~\tnote{1}  &0.34  &0.32 &0.41 &\bf{0.07}  \\ \hline
            DAER~\tnote{1}  &0.092 &\bf{0.032} &0.033 &\bf{0.032}   \\ \hline
            AEE-P~(m)~\tnote{2}    &20.49 &17.53 &17.33 &\bf{10.47}  \\ \hline
            AEE-V~(m/s)~\tnote{2}   &10.70 &10.66 &12.25 &\bf{3.12}  \\ \hline
            MOSPA~(m)~\tnote{3} & 26.61 &24.93 &26.76 & \bf{14.15}  \\ \hline
            TET~(s) &\bf{0.23} &5.71 &8.59 &14.27   \\ \hline
    \end{tabular}
      \begin{tablenotes}
    \footnotesize
    \item[1] MAER, DAER, AEE-P and AEE-V are averaged over all time steps.
    \item[2] AEE-P and AEE-V denote AEE  in position and velocity, respectively.
 \item[3] MOSPA is calculated with parameters $p = 2$ and $c = 100m$.
\end{tablenotes}
  \end{threeparttable}
\end{table}
Table~\ref{ComPd} provides an average performance comparison of MP-MMTT, IMMJPDA, IMMHMHT, and MMGLMB.
In the aspect of target detection performance evaluated by NVT and NFT,
the number of false tracks of MP-MMTT is much less than those of IMMJPDA, IMMHMHT, and MMGLMB although all the four algorithms can detect targets successfully.
In the aspect of target tracking performance, IMMJPDA, IMMHMHT, and MP-MMTT have comparable performance on persistent target tracking capability~(evaluated by TPD and ANBT).
The target tracking accuracy~(evaluated by AEE-P and AEE-V) of MP-MMTT is better than those of the other three algorithms.
This is because MP-MMTT adopts a batch processing methodology and a smooth mechanism where multiple scan measurements are integrated to improve detection and tracking performance.
Because of the batch and iterative processing structure, MP-MMTT makes a trade-off between estimation accuracy and computational cost.
The computational cost of MP-MMTT is the largest. Through the closed-loop structure where more accurate state estimates are used to identify data association and target motion model, MP-MMTT outperforms the other three algorithms in terms of MAER and OSPA, and has the comparable performance on DAER with IMMHMHT and MMGLMB.

To compare the computational cost of the four algorithms as the number of targets increases,
we vary the number of targets to be 8, 12, 16, respectively.
For each case, one-half ~(one subgroup) of the targets move in parallel along Y-direction and
the distance in X-direction between two neighboring targets is 100~m.
The rest~(the other subgroup) of the targets move in parallel along X-direction and the distance
in Y-direction between two neighboring targets is 100~m as well.
The two subgroups of targets cross and maneuver in the same way as the four targets.
Fig.~\ref{Time-num} shows the running time of the four algorithms when $N_T = 4, 8, 12, 16$.
It is observed that for the different number of targets, IMMHMHT has the lowest computational cost and IMMJPDA has the highest computational cost, while the computational cost of MP-MMTT and MMGLMB are comparable.
Note that when $N_T \geq 12$, IMMJPDA is not able to output results in a reasonable amount of time.
The computational cost comparison with different clutter densities $\lambda = 10^{-4}, 2 \times 10^{-4}, 4 \times 10^{-4}, 6 \times 10^{-4}, 8 \times 10^{-4}, 10^{-3}$ is shown as in Fig.~\ref{Time-pf}.
IMMHMHT and IMMJPDA are not computationally feasible for heavy clutter scenarios since their computational complexity scale exponentially in the number of~(valid) measurements, while MMGLMB has a lower computational cost than MP-MMTT.
Note that, however, MMGLMB improves the computational efficiency by assuming that the target birth intensity is known as a priori, which means the prior information of the region of interest is required.
Lack of such prior information will make the computational cost of MMGLMB increase dramatically.

More comparison results in the term of MOSPA for different values of clutter density, detection probability, spatial distance and number of targets are shown in Figs.~\ref{OSPA-pf}-\ref{OSPA-number}, respectively.
The MOSPA of MP-MMTT is smaller than those of other algorithms in all of these cases. Meanwhile,
the dotted square~(black) line represents the MOSPA performance curve of MP-MMTT without iteration~($r = 0$). Comparing it with the solid star~(red) line, which is the MOSPA performance curve of MP-MMTT when $r=3$, it is concluded that the closed-loop structure of MP-MMTT is indeed helpful to improve tracking performance.
The dash-dot diamond~(cyan) line represents the MOSPA curve of MP-MMTT with real-time outputs.
It is seen that, with the cost of delay outputs, the tracking performance of MP-MMTT is improved by adopting the smoothing strategy.

\section{Conclusion}\label{sec:conclusion}
This paper studied the joint detection and tracking of multiple maneuvering targets, and provided a derivation of unified MP approach that performs the BP and MF approximation for the joint estimation of target kinematic state and visibility state, and the decision of data association and motion mode-model association.
The corresponding beliefs were calculated iteratively via the fixed-interval smoother, the forward-backward algorithm and the LBP.
With the cost of time delay output of tracks like IMMHMHT, the proposed MP-MMTT method outperforms IMMJPDA, IMMHMHT and MMGLMB in the aspects of both target detection and tracking. Meanwhile, MP-MMTT is more computationally effective than IMMJPDA and IMMHMHT in the scenario of multitarget tracking with heavy clutter.

\bibliographystyle{IEEEtran} %the bibliographystyle
\bibliography{IEEEabrv,references} %the references

\end{document}